\def\etal{\it et al. \rm}
\def\kms{km~s$^{-1} \,$}
\def\kmsmpc{km~s$^{-1}$~Mpc$^{-1}\,$}

\def\Mo{${\rm M_\odot}$}

\def\gsim{ \lower .75ex \hbox{$\sim$} \llap{\raise .27ex \hbox{$>$}} }
\def\lsim{ \lower .75ex \hbox{$\sim$} \llap{\raise .27ex \hbox{$<$}} }
\def\spose#1{\hbox to 0pt{#1\hss}}
\def\simlt{\mathrel{\spose{\lower 3pt\hbox{$\mathchar"218$}}
     \raise 2.0pt\hbox{$\mathchar"13C$}}}
\def\simgt{\mathrel{\spose{\lower 3pt\hbox{$\mathchar"218$}}
     \raise 2.0pt\hbox{$\mathchar"13E$}}}
\def\gsim{ \lower .75ex \hbox{$\sim$} \llap{\raise .27ex \hbox{$>$}} }
\def\lsim{ \lower .75ex \hbox{$\sim$} \llap{\raise .27ex \hbox{$<$}} }
\def\simprop{ \lower .75ex \hbox{$\sim$} \llap{\raise .27ex \hbox{$\propto$}} }
\documentstyle[emulateapj,apjfonts,psfig,rotating]{article}

\lefthead{Frenk \etal}
\righthead{The Santa Barbara cluster comparison project.}

\begin{document}

%

\title{The Santa Barbara cluster comparison project: 
a comparison of cosmological hydrodynamics solutions}

\author{C. S. Frenk\altaffilmark{1}, S. D. M. White\altaffilmark{2}, 
P. Bode\altaffilmark{3}, J. R. Bond\altaffilmark{4},
G. L. Bryan\altaffilmark{5}, 
R. Cen\altaffilmark{6}, H.~M.~P.~Couchman\altaffilmark{7}, 
A. E. Evrard\altaffilmark{8}, N. Gnedin\altaffilmark{9},
A. Jenkins\altaffilmark{1}, 
A. M. Khokhlov\altaffilmark{10},
A.~Klypin\altaffilmark{11},
J. F. Navarro\altaffilmark{12},
M. L. Norman\altaffilmark{13,14}, 
J. P. Ostriker\altaffilmark{6}, 
J. M. Owen \altaffilmark{15,16},
F. R. Pearce\altaffilmark{1}, U.-L. Pen\altaffilmark{17}, 
M.~Steinmetz\altaffilmark{18}, P. A. Thomas\altaffilmark{19}, 
J. V. Villumsen\altaffilmark{2}, J. W. Wadsley\altaffilmark{4}, 
M. S. Warren\altaffilmark{20}, G. Xu\altaffilmark{21}, 
G. Yepes\altaffilmark{22} 
}

\affil{}

\altaffiltext{1}{Physics Dept., University of Durham, DH1 3LE, England}
\altaffiltext{2}{MaxPlanck Inst. f\"{u}r Astrophysik, Karl Schwarzschild 
Strasse 1,  D-85740 Garching bei Munchen, Germany}
\altaffiltext{3}{Dept. Physics and Astronomy, University of Pennsylvania, 
33rd and Walnut Street, Philadelphia, PA  19104-6396   USA}
\altaffiltext{4}{Canadian Inst. for Theoretical Astrophysics and Dept. of
Astronomy, University of Toronto, 60 George St, Toronto ON M5S 3H8, Canada}
\altaffiltext{5}{Physics Dept., MIT, Cambridge, MA 02139, USA}
\altaffiltext{6}{Princeton University Obs., NJ 08544, USA}
\altaffiltext{7}{Dept of Physics and Astronomy, University of Western Ontario, 
London, Ontario N6A 3K7, Canada}
\altaffiltext{8}{Dept. of Physics, University of Michigan, 
Ann Arbor, MI 48109-1120 USA}
\altaffiltext{9}{Astronomy Dept., UC Berkeley, CA 94720, USA}
\altaffiltext{10}{Lab. for Computational Physics and Fluid Dynamics, Code
  6404, Naval Research Laboratory, Washington,  DC, 20375, USA} 
\altaffiltext{11}{Department of Astronomy, New Mexico State University, Las
  Cruces, NM 88001, USA}
\altaffiltext{12} {University of Victoria, Dept of Physics \& Astronomy,
Victoria, BC, V8W 2P6, Canada}
\altaffiltext{13} {Lab. for Computational Astrophysics, NCSA, 
University of Illinois at Urbana-Champaign, 405 N. Mathews Ave., Urbana, 
IL 61801, USA}
\altaffiltext{14}{Astronomy Dept., University of Illinois at Urbana-Champaign,
1002 West Green Street, Urbana, IL 61801, USA}
\altaffiltext{15}{Dept. of Astronomy, Ohio State University, Columbus, OH 
43210, USA}
\altaffiltext{16}{Lawrence Livermore National Laboratory,
L-312, Livermore, CA  94550, USA}
\altaffiltext{17}{Harvard Smithsonian CfA, 60 Garden St, Cambridge, MA
02138, USA} 
\altaffiltext{18}{Steward Obs., University of Arizona, Tucson, AZ 85721, USA}
\altaffiltext{19}{CPES, University of Sussex, Falmer, Brighton BN1 9QH}
\altaffiltext{20}{Theoretical Astrophysics, T-6, Mail Stop B288, 
Los Alamos National Lab., Los Alamos, NM 87545, USA}
\altaffiltext{21}{Board of Studies in Astronomy and Astrophysics, UC Santa
Cruz, CA 95064, USA} 
\altaffiltext{22}{Depto. de F\'{\i}sica Te\'orica C-XI, Universidad
Aut\'onoma de Madrid, Cantoblanco 28049, Madrid, Spain}

\setcounter{footnote}{1}

\normalsize
\bigskip

\begin{abstract}
We have simulated the formation of an X-ray cluster in a cold dark matter
universe using 12 different codes. The codes span the range of numerical
techniques and implementations currently in use, including SPH and grid
methods with fixed, deformable or multilevel meshes. The goal of this
comparison is to assess the reliability of cosmological gas dynamical
simulations of clusters in the simplest astrophysically relevant case, that
in which the gas is assumed to be non-radiative. We compare images of the
cluster at different epochs, global properties such as mass, temperature
and X-ray luminosity, and radial profiles of various dynamical and
thermodynamical quantities. On the whole, the agreement among the various
simulations is gratifying although a number of discrepancies
exist. Agreement is best for properties of the dark matter and worst for
the total X-ray luminosity.  Even in this case, simulations that adequately
resolve the core radius of the gas distribution predict total X-ray
luminosities that agree to within a factor of two. Other quantities are
reproduced to much higher accuracy. For example, the temperature and gas
mass fraction within the virial radius agree to about 10\%, and the ratio
of specific kinetic to thermal energies of the gas agree to about
5\%. Various factors contribute to the spread in calculated cluster
properties, including differences in the internal timing of the
simulations.  Based on the overall consistency of results, we discuss a
number of general properties of the cluster we have modelled.
\end{abstract} 

\keywords{cosmology: theory --- dark matter --- galaxies: clusters ---
large-scale structure of universe }

\section{Introduction}

Computer simulations have played a central role in modern cosmology. Two
decades after the first cosmological simulations were performed, this
technique is firmly established as the main theoretical tool for studying
the non-linear phases of the evolution of cosmic structure and for testing
theories of the early universe against observational data.

The first generation of cosmological simulations employed N-body techniques
to follow the clustering evolution of a dissipationless dark matter
component. This approach proved powerful enough to reject the idea that the
dark matter consists of massive neutrinos and to establish the viability of
the alternative hypothesis that the dark matter is made up of cold
collisionless particles. In the last decade, N-body techniques have been
further refined and applied to a wide range of cosmological
problems. N-body simulations are now sufficiently well understood that the
validity of analytic approximations is often gauged by reference to
simulation results.

The main limitation of N-body techniques is, of course, that they are
relevant only to the evolution of dark matter. In order to model the
visible universe, it is necessary to include additional physical
processes. First, it is necessary to model a gas component that is
gravitationally coupled to the dark matter.  In the simplest case, the gas
may be assumed to be non-radiative. Although still highly simplified, this
case has immediate applications in the study of the hot plasma in galaxy
clusters. At the next level of complexity, heating and cooling processes
must be included. This is required, for example, to investigate the
physical properties of the gas clouds responsible for QSO absorption
lines. Additional processes, such as star formation and the associated
feedback of energy and mass, are necessary to model galaxy formation.

Since the late 1980s a variety of techniques have been developed to
simulate gas dynamics and related processes in a cosmological context. In
part inspired by the success of the N-body program, the first gas dynamical
techniques were based on a particle representation of Lagrangian gas
elements using the Smooth Particle Hydrodynamics (SPH) technique (Lucy
1977, Gingold \& Monaghan 1977, Evrard 1988). Soon thereafter, fixed-mesh
Eulerian methods were adapted (Cen \etal 1990, Cen 1992) and, more
recently, Eulerian methods with submeshing (Bryan \& Norman 1995) or
deformable moving meshes (Gnedin 1995, Pen 1995, 1998) have been developed,
as well as extensions of the SPH technique (Shapiro \etal 1996). These
codes are actively being applied to a variety of cosmological problems,
ranging from the formation of individual galaxies and galaxy clusters to
the evolution of Lyman-$\alpha$ forest clouds and the large-scale galaxy
distribution.

Because of the inherent complexity of gas dynamics in a cosmological
context, such simulations are more difficult to validate than N-body
simulations. Standard test cases with known analytical solutions (such as
shock tubes) are far removed from the conditions prevailing in cosmological
situations where the gas is coupled to dark matter and this, in turn,
evolves through a hierarchy of mergers.  The closest analogue to a
realistic cosmological problem is Bertschinger's (1985) solution for the
collapse of a spherical cluster. Although this model provides a useful test
of numerical hydrodynamics implementations, it ignores the merging
processes that are a dominant aspect of the formation of realistic
clusters. In general, the strongly non-linear and asymmetric nature of
gravitational evolution in a cosmological context differs greatly from the
regime that can be studied analytically or in laboratory experiments.

In this paper we carry out an exercise intended as a step towards assessing
the reliability of current numerical studies of cosmological gas dynamics.
We address one of the simplest astrophysically relevant problems, the
formation of a large cluster in a hierarchical cold dark matter (CDM)
model, using a variety of codes that span the entire range of numerical
techniques in use today. The cluster problem is relatively simple because,
except in the inner parts, the cooling time of the gas exceeds the age of
the Universe and so, to a good approximation, the gas may be treated as
non-radiative.

The aim of this exercise is to assess the extent to which existing
modelling techniques give consistent and reproducible results in a
realistic astrophysical application. Our comparison is, by design, quite
general. We simply specify the initial conditions for the formation of a
cluster and let different simulators approach the problem in the manner
they regard as most appropriate. Our comparison therefore encompasses not
only the hydrodynamics simulation techniques themselves, but also
individual choices of boundary conditions, resolution, internal variables
such as the integration timesteps, and even the definition of cluster
center. We are therefore able to address issues such as the reproducibility
of the X-ray luminosity and surface brightness of simulated clusters. It is
not our intention to test the accuracy of any individual code: all 
the codes used in this paper have already been extensively tested
against known analytic solutions. An earlier comparison of a subset of the
techniques considered here was presented by Kang \etal (1994b). These
authors simulated a large cosmological volume and focussed on statistical
properties of the large-scale structure, rather than on the non-linear
properties of an individual cluster that concern us here.

This project was initiated as part of the activities of the program on
``Cosmic radiation backgrounds and the formation of galaxies'' which took
place at the Institute for Theoretical Physics in Santa Barbara in
1995. Most active groups in the field of cosmological hydrodynamics
simulations agreed to participate. Initial conditions for the cluster
simulation were set up as described in Section~2. The codes used in the
comparison are briefly described in Section~3. Participants were asked to
analyze their results with a suite of predefined diagnostics, including
images, global properties such as mass and X-ray luminosity, and radial
profiles of the dark matter density, gas density, temperature, etc. The
images and a comparison of quantitative results are given in Section~4. Our
paper concludes in Section~5 with a summary and discussion of results,
including some general conclusions regarding the properties of the simulated
cluster. 

\section{The simulation}

We simulated the formation of a galaxy cluster in a flat CDM universe. The
initial fluctuation spectrum was taken to have an asymptotic spectral
index, $n=1$, and shape parameter, $\Gamma=0.25$, the value suggested by
observations of large-scale structure (eg. Efstathiou, Bond \& White
1992). The cosmological parameters assumed were: mean density, $\Omega=1$;
Hubble constant, $H_0=50$\kmsmpc; present-day linear rms mass fluctuations
in spherical top hat spheres of radius 16 Mpc, $\sigma_8=0.9$; and baryon
density (in units of the critical density), $\Omega_b=0.1$.

\subsection{ Initial conditions} 

Initial conditions were laid down using the constrained Gaussian random
field algorithm of Hoffman \& Ribak (1991). The cluster perturbation was
chosen to correspond to a $3\sigma$ peak of the density field smoothed with
a Gaussian filter of radius $r_0 = 10$ Mpc [in $\exp(-0.5(r/r_0)^2)$]. The
perturbation was centered on a cubic region of side $L=64$ Mpc.  We used
the fit to the CDM transfer function given by eqn~(G3) of Bardeen \etal
(1986) and recommended a starting epoch of $z=20$.

To offer flexibility, the initial conditions were generated at very high
resolution. Two alternative forms were supplied:

\begin{itemize}
\item[(i)] The dimensionless linear $\delta \rho/\rho$ field (normalized to the
present), tabulated on a $256^3$ cubic mesh.
\item[(ii)] The linear theory displacements for $256^3$ points on a cubic 
mesh.
\end{itemize}
\noindent The initial conditions were generated by Shaun Cole. These are
publically available on the Internet at
http://star-www.dur.ac.uk/~csf/clusdata/ or by request from CSF.

\subsection{Simulation diagnostics} 

The simulation data were output at redshifts $z=$8, 4, 2, 1, 0.5, and 0 and
the following properties were calculated:

\medskip
\noindent{\it Images.}
\medskip

At each epoch, 2-D images were generated of various quantities in the
central (32 Mpc)$^3$ comoving volume. The quantity of interest was
projected along the $z$-axis, smoothed as specified below, and tabulated on
a 1024$^2$ mesh. Two smoothings were employed:

(i) A Gaussian smoothing with kernel, $\exp(-0.5(r/r_0)^2)$, at a fixed
resolution of $r_0=250$ kpc (comoving).

(ii) A smoothing of choice, as determined by each simulator.

\noindent The first set of images was used for a uniform comparison of all the
models, while the second set was supplied in order to display the results
of each technique in the best possible light.

Images of the following quantities were made:

(i) Projected dark matter density (in \Mo Mpc$^{-2}$)

(ii) Projected gas density (in \Mo Mpc$^{-2}$)

(iii) X-ray surface brightness, $\int {\cal L}_X dl$ (X-ray emissivities
per unit volume were calculated as ${\cal L}_X=\rho^2 T^{1/2}$, with $\rho$
in \Mo Mpc$^{-3}$ and $T$ in K.)

(iii) Emission-weighted temperature $(\int {\cal L}_X T dl/\int {\cal L}_X
dl)$.


\medskip
\noindent{\it Global properties.}
\medskip 

We defined the cluster to be the mass contained inside a sphere of radius,
$r_{200}$, such that the mean interior overdensity is 200.  The following
global properties of the cluster were then computed at $z=0$:

(i) The value of $r_{200}$

(ii) $M_{\rm dm}:$ \ total dark matter mass (in \Mo)

(iii) $V_{\rm dm}:$ \ {\it rms} velocity of dark matter particles (in km
                      s$^{-1}$)

(iv) $M_{\rm gas}:$ \ total gas mass (in \Mo )

(v) $\overline T:$ \quad \ mean (mass weighted) temperature (in K)

(vi) $U:$ \quad \ total bulk kinetic energy of the gas (in ergs)

(vii) $L_{\rm tot}= \int_0^{r_{200}} {\cal L}_X dV$ (${\cal L}_X$ units as
                  above; $V$ in Mpc$^{3}$)

(viii) ${\cal I} = \sum_i m_i {\bf x}_i {\bf x}_i /\sum_i m_i$: inertia
                tensor for dark matter and gas (in Mpc$^2$).

\medskip
\noindent {\it Radial profiles.}
\medskip
 
In 15 spherical shells of logarithmic width 0.2 dex and inner radii $10
{\rm kpc} \le r < 10 {\rm Mpc}$, the following quantities were obtained:

(i) $\rho_{\rm dm}(r):$ dark matter density profile (in \Mo Mpc$^{-3}$)

(ii) $\sigma_{\rm dm}(r):$ dark matter velocity dispersion profile (in km
                            s$^{-1}$)

(ii) $\rho_{\rm gas}(r):$ gas density profile (in \Mo Mpc$^{-3}$)

(iv) $T(r):$ \quad mass-weighted gas temperature profile (in K)

(v) ${\cal L}_X(r):$ ``X-ray luminosity" profile calculated as the total
                       luminosity in each bin, divided by its volume (in
                       units as above).

Each simulator was provided with the two initial conditions files and the
list of required diagnostics (in a prespecified format). Everything else
was left to the discretion of each simulator including, for example, the
definition of the cluster center. All data were sent directly to the
organizers (CSF and SDMW), and participants were strongly discouraged from
private intercomparison of results. A surprising number of iterations was
required to obtain consistent outputs in a single set of units and formats.

Wadsley joined the project after the original deadline had expired and the
first set of results was known. Discrepancies in a preliminary comparison
of results led Gnedin to revise his code and resubmit a new simulation.
Bryan's stated spatial resolution was changed from 80 to 30 kpc after
preliminary comparison suggested that he had been too pessimistic in
stating his resolution.

\section{ The codes} 

The numerical codes used for this project employ a variety of techniques to
solve the evolution equations for a two component fluid of dark matter and
non-radiative gas coupled by gravity. In what follows, each code is
identified by the name of the author who was primarily responsible for
carrying out each simulation. The codes are of two general types: SPH and
grid-based. SPH simulations were carried out by Couchman, Evrard, Jenkins,
Navarro, Owen, Steinmetz and Wadsley. The simulations by Couchman and
Jenkins used the same basic code (HYDRA), the serial version in the former
case and a parallel version in the latter. (These two simulations were done
independently and used different numbers of particles and different values
for the simulation parameters: gravitational softening, smoothing length,
timestep, etc.) Owen's code differs from the others in the use of an
anisotropic SPH kernel. The grid-based methods employ either a single,
fixed mesh (Cen, Yepes), a 2-level multi-mesh (Bryan) or a deformable mesh
(Gnedin, Pen).  Warren carried out a high resolution simulation of the
evolution of the dark matter only.

A brief description of each code follows, together with references where
the reader may find a fuller discussion of techniques and the tests to
which each code has been subjected. Details of each simulation are given in
Table~1.

\subsection{SPH codes} 


\subsubsection{Couchman \& Thomas -- Hydra (Adaptive P$^3$M-SPH)}

Hydra is functionally equivalent to the standard
particle-particle-particle-mesh, N-body-SPH (P$^3$M--SPH) implementation,
but with the automatic placement of a hierarchy of refined meshes in
regions of high particle density. This avoids the dramatic performance
degradation caused by the direct summation (PP) component of standard P$^3$M
codes under heavy particle clustering. In the present simulation, at a
redshift of $z=0.5$, the cpu time per step had increased by a factor of 4.5
from the essentially uniform initial conditions at $z=49$, and remained at
this level to the end of the simulation. A maximum of four levels of mesh
refinment was chosen by the code.

The code automatically chooses a global timestep to ensure accurate time
integration. This value is determined by the maximum instantaneous values
of particle velocities and accelerations of both gas and dark matter
particles. An optimal low-order integration scheme is used for advancing
particle positions and velocities. Full details of the code are available
in Couchman, Thomas and Pearce~(1995) and the source code may be found in
Couchman, Pearce and Thomas (1996).

The supplied initial displacement field was degraded to the resolution
used, $64^3$ dark matter particles, simply by sampling every fourth
position in each dimension. This rather crude method of resampling,
although simple, has the disadvantage of introducing noise into the
perturbed particle distribution above the effective Nyquist frequency of
the $64^3$ particles. Dark matter particles were displaced from a uniformly
spaced grid and gas particles were placed on top of the dark matter
particles. Particle displacements were scaled to correspond to a start
redshift of 49. The center of the final cluster was identified with the
density peak within $r_{200}$.

\subsubsection{Jenkins \& Pearce  -- Parallel Hydra} 

This simulation used a parallel version of Hydra, the code just described
in 3.1.1, whose distinguishing feature is its ability to place high
resolution meshes recursively around clustered regions. The SPH calculation
used an M4 spline kernel containing an average of 32 particles. Details of
the serial version of this code may be found in Couchman, Thomas and
Pearce~(1995), while details of the parallel implementation may be found in
Pearce \etal (1995) and Pearce \& Couchman (1997).

The simulation was carried out on a Cray-T3D. Initial conditions were laid
down by perturbing 128$^3$ particles distributed in a ``glass''
configuration. This was generated in the manner described by White (1996),
ie. by evolving a Poissonian distribution of points, with the sign of
gravity reversed, over many thousands of expansion factors. To optimize the
resolution in the region of interest, the computational volume was divided
into two parts, a high resolution spherical region containing $1/4$ of the
volume and centered on the location of the constrained peak, and a coarsely
sampled exterior region. Dark matter and gas particles (initially
coincident) were placed in the high resolution region, and dark matter
particles only in the exterior region. The coarse sampling was achieved by
smoothing the distribution with a nearest grid point (NGP) assignment on a
$64^3$ mesh, so that, on average, each particle was 8 times more massive
than particles in the high resolution region. This procedure reduced the
particle number from 2097152 dark matter particles to 1247217 of both
species, about one million of which lay in the high resolution region. The
initial particle positions were set up at $z=20$ using a trilinear
interpolation of the displacement field in the 8 mesh points surrounding
each particle. Velocities were assigned from the Zel'dovich approximation.
The center of the final cluster was defined to be the position of the
particle with the lowest gravitational potential.

Since this and Couchman's simulation were carried out with the same code,
one in parallel and the other in serial mode, any differences in the
results must be due to differences in the initial conditions or the choice
of integration parameters. We have checked that running Couchman's initial
conditions in parallel mode does not alter his results in any significant
way.

\subsubsection{Evrard -- P$^3$M-SPH} 

The P$^3$M-SPH code combines the P$^3$M code of Efstathiou \& Eastwood
(1981) with an adaptive kernel SPH scheme, as described in Evrard (1988).
The simulation for this study employed a two--level mass hierarchy, with a
high resolution ($64^3$ effective) inner zone of both dark matter and gas
surrounded by dark matter at low resolution ($32^3$).  The $N^3$ mesh data
were generated by NGP subsampling of the original $256^3$ displacement
field.  The mapping of the high resolution zone was determined by a low
resolution ($32^3$) N--body simulation; particles within a final density
contrast of 6 centered on the group in this run define a Lagrangian mask
used to generate the two--level initial conditions of the full run.  Masked
locations in the $32^3$ subsampled field were locally ``exploded'' to a
factor 2 higher linear resolution, generating an effective $64^3$
resolution within the non--linear parts of the cluster.  This procedure
assures no contamination of low resolution particles within the cluster in
the production run.  The run used 30456 particles for each high resolution
component (gas and dark matter) and 28961 particles at low resolution, with
a $128^3$ Fourier mesh for the long--range gravity.  The center of the
final cluster was defined by the most bound dark matter particle.
    
The number of interacting neighbors within the smoothing kernel controls
the hydrodynamics resolution of the calculation.  This parameter was set so
that approximately 168 particles lie within a sphere of radius $2h$ around
any particle.  As discussed by Owen and Villumsen below, the value of this
parameter varies considerably among experiments in the literature.  The
value employed here is larger than ``typical'' values and reflects a desire
to minimize the Poisson noise inherent in the kernel summations required
for calculation of the density and pressure gradient terms.

The computation was performed on a local HP workstation.  The modest memory
and CPU requirements of this calculation reflect its nature as closer to
``everyday'' than ``state--of--the--art''.  It is representative of the
type of runs used in ensembles to investigate statistical aspects of the
cluster population (e{.}g{.}  Mohr \& Evrard 1997).

\subsubsection{Navarro -- Grape+SPH} 

The code used was the N-body/SPH code described by Navarro \& White
(1993), adapted to compute gravitational accelerations using a GRAPE-3
board. The neighbor lists needed for the SPH computations are also
retrieved from the GRAPE and processed in the front-end workstation. The
implementation of these modifications is straightforward and very similar
to that described by Steinmetz (1996), where the reader may find far more
details.

The initial conditions were realized by perturbing a cubic grid of
particles with the displacement field made available with the initial
conditions package.  The system was divided in two zones, an inner cube of
size 38 Mpc which was filled with $40^3$ dark matter and $40^3$ gas
particles, surrounded by a sphere of diameter $64$ Mpc. The region outside
the inner cube was filled with $\sim 5,000$ low-resolution dark matter
particles of radially increasing mass.  Initially, gas and dark matter
particles were placed on top of each other and were given the same
velocities, computed using the Zel'dovich approximation.  The final cluster
center was calculated using a concentric sphere method that isolates the
highest density peak iteratively by computing the center of mass of a
sphere and successively removing the outermost particle, until only about
100 particles are left.

The simulation was run on the SPARC10/GRAPE-3 system at Edinburgh
University.

\subsubsection{Steinmetz -- GrapeSPH} 

This simulation was performed using GrapeSPH (Steinmetz 1996), a direct
summation hybrid N-body/SPH code especially designed to take advantage of
the hardware N-body integrator GRAPE (Sugimoto \etal 1990). It is highly
adaptive in space and time through the use of individual particle timesteps
and individual smoothing lengths. Details of the code such as the adaptive
smoothing length or the multiple time stepping procedure, are presented in
Steinmetz \& M\"uller (1993) and in Steinmetz (1996).

The simulation used a multi-mass technique similar to that described by
Porter (1985). Firstly, a low resolution ($32^3$ particles) P$^3$M
simulation of the full periodic volume was performed. The initial
conditions were drawn from the distribution supplied by averaging positions
and velocities in cubes of $8^3$ particles. At $z=0$ the cluster which
formed near the center was identified and its virial radius, $r_{200}$,
determined. Particles within $r_{200}$ were marked and traced back to
redshift $z=20$. A sphere was then drawn containing all these particles --
the high resolution region. Particles within that sphere were replaced by
the corresponding particles drawn from higher resolution initial
conditions. Thus, the particle number in the high resolution region was
increased by factors of 8-64.  Particles outside the high resolution region
were combined into larger mass nodes using the tree-pruning technique
described in Porter (1985). The mass of a particle thus increases
logarithmically with distance from the central sphere.  Starting from these
initial conditions, the full simulation was performed, with gas dynamics
followed only in the high resolution region. The center of the final
cluster was defined as the center of mass of the smallest (radius 125 kpc)
of a series of 7 concentric spheres of progressively decreasing
radius. (This agreed to about 5\% with the minimum of the gravitational
potential.)

In the hardware integrator GRAPE, the interparticle force is hardwired to
obey a Plummer force-law. Thus, periodic boundary conditions cannot easily
be realized (for a more recent development, see Huss, Jain \& Steinmetz
1998). Because of the tree-pruning, however, the CPU time scales only
logarithmically with box size for a given numerical resolution.  A typical
application thus starts from a very large simulation sphere assuming vacuum
boundaries. Since the computational box supplied for this cluster
simulation was relatively small, effects due to the finite box size cannot
be excluded and this may also affect the comparison with grid based
methods.  In order to minimize the effects of finite box size and vacuum
boundary conditions, the simulation strategy was slightly modified. Tree
pruning was not applied to the original box, but to an enlarged box
including the 26 neighboring periodic replicas of the original. Hence,
vacuum boundaries apply to a surface of radius $r=1.5\,l_{\rm box}$, rather
than $r=0.5\,l_{\rm box}$.

A variety of simulations with differing numerical resolution, particle
numbers and size of the high resolution region were performed. Results from
one simulation only have been included in this paper. This probably
reflects the best compromise between resolution and computational cost.  In
this simulation, which took about 28 hours of CPU and 22 MBytes of memory,
$\sim 15000$ gas and dark matter particles ended up within the virial
radius of the cluster at redshift $z=0$, a resolution similar to that
achieved by Evrard, Navarro, Couchman and Wadsely.  The largest simulation
carried out had the same number of gas particles but 8 times as many dark
matter particles.  This run consumed a total CPU time of 254 hours and
required 45 MBytes of memory.

\subsubsection{Wadsley \& Bond -- P$^3$MG--SPH}

The Wadsley and Bond (1997) $P^3MG$--$SPH$ code used in this cluster
comparison combines SPH for the hydrodynamics with an iterative multigrid
scheme to solve for the non-periodic gravitational potential with a
particle-particle correction for subgrid forces. A recursive linked list is
used to locate neighbor particles for SPH. At each timestep, a multipole
expansion is used to obtain the gravitational potential boundary conditions on
a $128^3$ grid.  The multigrid technique is quite competitive with Fast
Fourier Transform methods in speed and can more efficiently treat
non-periodic configurations, for which the $P^3MG$--$SPH$ code is designed. 
It is typically used to compute highly active inner regions at high
resolution, with large scale tides treated using a sequence of
progressively lower resolution spherical shells in the initial
conditions. The force is augmented by a measured external tidal field
evolved using linear theory.

This cluster had a Gaussian filter scale for the peak which was too large
for the periodic box size to treat the tidal environment adequately ({\it
i.e.}, the cluster was apodized). To a spherical high resolution region of
radius 25 Mpc, we could only add a single lower resolution shell extending
to 32 Mpc in radius. The 25 Mpc choice was based on the region the
peak-patch theory (Bond and Myers 1996) suggests would have collapsed. The
low resolution particles had 8 times the mass of the high resolution ones.
The self consistent linearly-evolved external shear was also applied to the
entire region. There were 74127 gas and 74127 dark particles used in the
simulation. The initial $256^3$ displacement field was sampled at every
fourth lattice site to transfer onto the computational grid for the high
resolution region. A similar transfer was done for the low resolution
region, with slight smoothing added.  Couchman used the same one-in-four
transfer method, probably accounting for the similarities with the Wadsley
and Bond result, especially with regard to timing. Discrepancies may be
due, at least in part, to his use of periodic boundary conditions. The
center of the final cluster was taken to be the center of mass of the
largest group in the simulation identified with a standard
friends-of-friends group finder.

The computation was run on a Dec-Alpha EV5 and required 119 CPU hours and
100 Mb of memory. The current version of this code has a significantly
accelerated particle-particle section, using tree techniques (the PP
section of the old gravity solver slowed by a factor of 10 by $z=0$).  The
same computation now takes 33.6 CPU hours and 50 Mb of memory.

\subsubsection{Owen \& Villumsen -- Adaptive SPH}

This simulation was performed with a variant of the SPH method called
Adaptive Smoothed Particle Hydrodynamics, or ASPH.  ASPH generalizes the
isotropic sampling of SPH by associating an individual, ellipsoidal
interpolation kernel with each ASPH node, the size, shape, and orientation
of which is evolved using the local deformation tensor $\partial
v_i/\partial x_j$.  The goal of the algorithm is to maintain a constant
number of particles per smoothing length in all directions at all times for
each ASPH particle.  This anisotropic sampling allows the ASPH resolution
scale to better adapt to the local flow of material as compared with the
isotropic sampling of traditional SPH, thereby maximizing the resulting
spatial resolution for a given number of particles.  Another way of stating
this is to say that in the frame defined by the kernel, the distribution is
locally isotropic.

The ASPH formalism is meaningful only when particles are treated not as
particles but as moving centers of interpolation.  The effects of momentum
non-conservation are minimized so long as internal consistency in the ASPH
kernel field is maintained, a condition we enforce by renormalizing the
ASPH kernels periodically.  The prescription for this renormalization, the
ASPH algorithm, and the code used here are described in detail in Owen et
al.\ (1997), and an earlier discussion of ASPH may be found in Shapiro et
al.\ (1996).

The ASPH interpolation between nodes is performed using the bi-cubic spline
interpolation kernel, which formally extends for two smoothing scales, $h$.
Beyond this point, the bi-cubic spline falls to zero, and therefore only
nodes within $2 h$ can interact with each other.  The local smoothing
scales were initialized such that there are roughly 2 nodes per smoothing
scale, so each node ``sees'' a radius of 4 nodes, or a total of $4/3 ~\pi
~4^3 \approx 268$ nodes.  Of course, since the bi-cubic spline weighting
falls to zero near the edge of this sampling volume, each node effectively
interacts with only about $4/3 ~\pi ~3^3 \approx 113$ neighbors.  While
this represents a much larger number of neighboring particles than most
contemporary SPH implementations (a more typical value is 1 particle per
$h$, yielding 32 particles in a volume of radius $2 h$) it appears that
keeping 1.5--2 particles per $h$ yields more reliable results for a wide
variety of hydrodynamical test problems (see also Balsara 1995). Evrard's
SPH simulation also used a a large number of interacting particles. The
disadvantage of this choice, of course, is that it decreases the effective
resolution.  An additional feature of the code is that it uses a compact,
higher order interpolation kernel for the artificial viscous interactions
in an effort to more closely confine the effects of the artificial
viscosity to shocked regions.  The gravitational interactions are evaluated
using a straight, single-level Particle-Mesh (PM) technique.

This experiment was performed using $32^3$ dark matter and $32^3$ ASPH
particles, and a $256^3$ PM grid for the gravity.  Since there are equal
numbers of ASPH and dark matter particles, each dark matter particle is 9
times as massive as an ASPH particle.  The initial conditions were
generated for $z=20$ by perturbing the $2 \times 32^3$ ASPH and dark matter
particles (initially exactly overlaying each other) from a cubical lattice
with trilinear interpolation based upon the supplied displacement field,
and assigning velocities using the Zel'dovich approximation. The center of
the final cluster was defined through an iterative approach, similar to
that used by Steinmetz. In this case, the radius of each successive sphere
was shrunk by a factor of 0.9 and the iterations were stopped when the
center of mass shifted by less than a small tolerance.

It is worth commenting on the fact that with only $32^3$ ASPH and dark
matter particles, this is by far the lowest resolution of the SPH
experiments presented in this paper.  This is primarily due to the fact
that at the time this experiment was performed our 3-D ASPH code had only
just been completed, and this was one of the first problems tackled with
that (then) highly experimental code.  Due to the lateness of our entry
into this project, we only had time to confirm that the experiment appeared
to have run successfully and submit that initial run.  The current version
of our 3-D ASPH code is competitive with other contemporary cosmological
hydrodynamical codes.  Despite these limitations, though, it is interesting
to compare the results of this simulation with the others in order to
quantify how well the regions which are resolved match the results of the
higher resolution models.

\subsection{Grid-based methods}

\subsubsection{Bryan \& Norman -- SAMR}

A newly-developed, structured, adaptive mesh refinement (SAMR) code was
used to perform this simulation.  This method was designed to provide
adaptive resolution while preserving the shock-capturing characteristics of
an Eulerian hydrodynamics scheme.  The code identifies regions requiring
higher resolution and places one or more finer sub-meshes over these areas
in order to better resolve their dynamics.  There is two-way communication
between a grid and its `child' meshes: boundary conditions go from coarse
to fine, while the improved solution on the finer mesh is used to update
the coarse `parent' grid.  The grid placement and movement is done
automatically and dynamically, so interesting features can be followed at
high resolution without interruption.  Since sub-grids can have
sub-sub-grids, this process is not limited to just two levels.  The control
algorithm for advancing the grid hierarchy is similar to that suggested by
Berger \& Colella (1989), and the equations of hydrodynamics are solved on
each grid with a version of the piecewise parabolic method (PPM) modified
for cosmology (Bryan et al. 1995).  Dark matter was modeled with particles,
and gravitational forces were computed via an adaptive particle-mesh
scheme.  Poisson's equation was solved on each mesh using the Fast Fourier
Transform.

The simulation was initialized at $z=30$ with two grids already in place.
The first is the root grid covering the entire 64 Mpc$^3$ domain with
$64^3$ cells.  The second grid is also $64^3$ cells but is only 32 Mpc on a
side and is centered on the cluster, yielding an initial cell size of 500
kpc (the initial conditions were provided at higher resolution but were
smoothed with a sharp k-space filter where appropriate).  The refinement
criteria was based on local density, so any cell with a baryon mass of $3.5
\times 10^9 M_\odot$ or more, was refined, but only within the central,
high-resolution region.  At $z=0.5$, the hierarchy consisted of more than
300 grids spread out over 7 levels of refinement.  Each level had twice the
resolution of the one above, producing, in very small regions, a cell size
of 8 comoving kiloparsecs.  The final cluster center adopted was the
cell-center of the cell with the highest baryonic density. The simulation
was carried out on four processors of an SGI Power Challenge at the
National Center for Supercomputing Applications (NCSA).

\subsubsection{Cen, Bode, Xu \& Ostriker  -- TVD}

This simulation employed a new shock-capturing Eulerian cosmological
hydrodynamics code based on Harten's Total Variation Diminishing (TVD)
scheme (Harten 1983), and described in Ryu \etal (1993).  The
original TVD scheme was improved by adding one additional variable
(entropy) and its evolution equation to the conventional hydrodynamics
equations.  This improvement eliminates otherwise large
artificial entropy generation in regions where the gas is not shocked.
Details of this treatment can be found in Ryu \etal (1993).  The
code is able to capture a strong shock within 1-2 cells and a sharp density
discontinuity within 3-5 cells.  Poisson's equation is solved using the
Fast Fourier Transform. The code is accurate in terms of global energy
conservation to about 1\%, as gauged by the Layzer-Irvine equation.

Initial conditions were laid down on a $512^3$ uniform grid, using the
particle positions, velocities and gas densities provided.  The initial gas
temperature was set to a low value, $113\;$K.  The total number of
particles was $256^3$ and the total number of fluid cells was $512^3$.  The
simulation started at $z=40$. The final cluster center was taken to be the
cell with the highest X-ray luminosity.

The simulation was run on an IBM SP2 at the Cornell Theory
Center. Sixty-four SP2 processors were used for the simulation for about 83
wallclock hours with 600 timesteps.  The code is well parallelized on the
SP2 (as well as on the Cray-T3E) and achieves an efficiency of about 50\%
on 64 SP2 processors (Bode \etal 1996).

\subsubsection{Pen -- Moving Mesh Hydrodynamics}

The simulations were all performed with an early version of the Moving Mesh
Hydrodynamics and N-body code (MMH for short; Pen 1995, 1998).  Its main
features are a full curvilinear TVD hydro code with a curvilinear PM N-body
code on a moving coordinate system.  The full Euler equations are solved
using characteristics in explicit flux-conservative form.  The curvilinear
coordinates used in the code are derivable from a gradient of the Cartesian
coordinate system.  If $x^i$ are the Cartesian coordinates, the curvilinear
coordinates are $\xi^i = x^i + \partial_\xi^i\phi({\bf \xi})$.  The
transformation is completely specified by the single potential field
$\phi({\bf \xi},t)$.  During the evolution any one constraint can be
satisfied by the grid.  In our case, we follow the mass field such that the
mass per unit grid cell remains approximately constant.  This gives all the
dynamic range advantanges of SPH combined with the speed and high
resolution of grid algorithms. For reasons of cpu economy (the
computational cost increases linearly with compression factor), this run
was constrained to compress by at most a factor of 10 in length, or a
factor of 1000 in density.

The potential deformation maintains a very regular grid structure in high
density regions.  The gravity and grid deformation equations are solved
using a hierarchical multigrid algorithm for linear elliptic equations.
These are solved in linear time, and are asymptotically faster than the FFT
gravity solver.  At the same time, adaptive dynamic resolution is
achieved. The gravitational softening of 45 kpc listed in Table~1 applies
to the central region of the cluster; the average softening is 450 kpc. The
final cluster center was identified with the minimum in the gravitational
potential.

The algorithmic cost per particle per timestep is very small, $\sim 300$
FLOP (floating point operations).  The cost for the grid deformation,
gravity and hydrodynamics adds to about 20k FLOP per grid cell per
timestep. If memory is available, we always use 8 particles per grid cell, at
which point the particles only account for a small portion of the
computation time.  This ensures that we are unlikely to encounter artifacts
due to 2-body relaxation.

The code runs in parallel on shared memory machines without load balancing
problems.  The simulation was carried out using a $128^3$ grid and $256^3$
particles.  Currently each timestep takes 60 seconds on a 16 processor
Origin 2000 at 195 Mhz.  The whole run takes 1600 time steps or about one
day.  At the time the actual simulations were performed, the best available
machine was a 75 Mhz R8k Power Challenge, where the run on 8 processors
took 60 hours.

Initial conditions were specified on an initially uniform grid.  The fluid
perturbation variables were set up on the grid, and particles displaced
using the Zel'dovich approximation. (At the time of writing, the code no
longer uses the Zel'dovich approximation, but instead varies the mass of
each particle.) The initial grid can now be adjusted to resolve
hierarchically any region of interest with arbitrary accuracy.  We expect
the new version to perform significantly better.

\subsubsection{Gnedin -- Smooth Lagrangian Hydrodynamics}

In this code, the hydrodynamical evolution of the gas is followed using the
Smooth Lagrangian Hydrodynamics or SLH method (Gnedin 1995), in which all
physical quantities are defined in quasi-Lagrangian space, $q^k$, and
Eulerian positions, $x^i$, are considered as dynamical variables.  The
imaginary mesh connecting Eulerian positions, $x^i$, thus moves with the
fluid until one of eigenvalues of the deformation tensor,
$A^i_k\equiv\partial x^i/\partial q^k$, becomes smaller than the predefined
softening parameter, $\lambda_*$. Then in the direction corresponding to
this eigenvalue the mesh gradually decelerates and progressively approaches
(but never fully reaches) the locally stationary mesh, until (and if) the
corresponding eigenvalue of the deformation tensor begins to increase. This
process of softening of the Lagrangian flow prevents severe mesh
distortions which can cause the stability and accuracy of a purely
Lagrangian code to deteriorate.  The gravitational force in the code is
computed using the P$^3$M method and is subject to the Gravitational
Consistency Condition as described in Gnedin \& Bertschinger (1996).

Initial conditions were set up by sampling the supplied fields on a $64^3$
mesh and using the Zel'dovich approximation to advance the dynamic
variables to $z=20$. The cluster center was defined using the DENMAX
algorithm.

\subsubsection{Yepes, Khokhlov \& Klypin  -- PM-FCT}

The code used for this simulation is a combination of an Eulerian
hydrodynamical code based on the Flux-Corrected-Trans\-p\-ort (FCT)
technique (Boris 1971, Boris \& Book 1973, 1976) and a standard
Particle-Mesh N-body code (Kates, Kotok and Klypin 1990). It uses the ``low
phase error algorithm'' whereby phase errors in convection are reduced on
the uniform grid to fourth order (Boris \& Book 1976, Oran \& Boris 1986).
This algorithm is applied to the hydrodynamics equations in one dimension.
At each timestep these are first integrated by FCT for a half-step to
evaluate time-centered fluxes; the FCT is then applied to a full timestep.

Multiple dimensions are treated through directional timestep splitting.  In
multiple dimensions, the code has overall second-order accuracy in regions
where the flow is continuous and provides a sharp, non-oscillating solution
near flow discontinuities. To avoid excessive temperature fluctuations at
shocks, the gas density is smoothed over the seven nearest nodes (one cell
in each direction) when estimating the temperature from the total energy,
velocity and density. This smoothing is done {\it only} for temperature
estimates.  Tests of the code and applications to cosmological problems may
be found in Klypin \etal (1992) and Yepes \etal (1995, 1996).

The code has been fully parallelized for various shared memory
platforms. The simulation reported here was performed on the CRAY-YMP at
CIEMAT (Spain) using 4 processors simultaneously. Due to memory
limitations, the supplied initial conditions were resampled from the
original $256^3$ grid onto a coarser grid with 160 cells and particles per
dimension. The initial particle positions were set up at $z=20$ using
cloud-in-cell interpolation of the original displacement field and
velocities were assigned by means of the Zel'dovich approximation.  The
cluster center was found iteratively from the center of mass of the
particle distribution in spheres of radius equal to 2 cells.

\subsection{Dark matter only}

\centerline{\it Warren -- Tree} 

This dark matter only simulation was carried out using a parallel treecode
(Warren and Salmon 1993, Warren and Salmon 1995) on 128 processors of the
512 processor Intel Delta at Caltech.  The algorithm computes the forces on
an arbitrary distribution of masses in a time which scales with the
particle number, $N$, as $N\log N$.  The accuracy of the force calculation
is analytically bounded, and can be adjusted via a user defined parameter.

Initial conditions were obtained by perturbing the masses of the particles
in proportion to the values of the supplied initial density field, starting
at a redshift of 63.  Growing mode velocities were assigned using the
Zel'dovich approximation.  The initial conditions were coarsened at radii
exceeding 24 Mpc by grouping cells 8 to 1, resulting in a total of
$5\,340\,952$ dark-matter particles in the simulation.  Periodic boundary
conditions were implemented by using the treecode to obtain forces from the
26 neighboring cube images, and an analytic treatment for the remainder.
The initial portion of the simulation (to $z=9$) was performed with a
comoving Plummer softening of 50 kpc, and a logarithmic timestep.  At $z =
9$, the softening was fixed at 5 kpc in physical coordinates, and a global
timestep of .005 Gyr was used, resulting in 2550 total timesteps for the
simulation.  The center was defined as the particle with the highest
density, smoothed with a spline kernel of width 20 kpc.

The upper limit for the fractional interaction force error was set to
0.005.  In the initial stages of the simulation, about 1200 interactions
per particle were computed.  Near the end, this had grown to about 2300
interactions per particle.  In terms of wall-clock time, this corresponded
to about 144 seconds per timestep initially, and 215 seconds per timestep
towards the end, representing a sustained throughput of roughly 1.5 Gflops
on the 128 processors.

\placetable{tbl-1}

\section{Results}

We first present a qualitative comparison of the results of the different
simulations using a selection of the images. We then discuss quantitative
results for the bulk properties and radial profiles of the clusters.

\subsection{Images} 

The images display projections of the following quantities in the inner
8~Mpc cube of each simulation:

(i) Dark matter density at $z=0$ (Figures~1 and~8) and $z=0.5$ (Figure~2),

(ii) Gas density at $z=0$ (Figure~3) and $z=0.5$ (Figure~4),

(iii) Gas temperature at $z=0$ (Figure~5) and $z=0.5$ (Figure~6),

(iv) X-ray luminosity at $z=0$ (Figure~7).

\noindent In Figures~1-6, the standard smoothing (250 kpc) was used,
whereas in Figures~7 and~8 the optimal smoothing chosen by each simulator
was used (see Table~1). The time elapsed between the two epochs shown in
the Figures is $6\times 10^9$ years, almost exactly half the dynamical time
of the final cluster (defined as $t_{dyn}=2\pi(r_v^3/{\rm G} M)^{1/2}$
where $r_v$ is the virial radius and $M$ the mass within it). Warren's dark
matter simulation is illustrated only in Figures~1, 2 and~8, in the bottom
right hand corner occupied in the remaining figures by Wadsley's
simulation, which was the last to be completed. Wadsley's dark matter
distribution has a very similar appearance to Couchman's.

\medskip
\noindent{\it Dark matter density}. 
\medskip

All simulations show a pleasing similarity in the overall appearance of the
projected dark matter density at the final epoch (Figures~1 and~8). The
size, shape and orientation of the main mass concentration are very similar
in all cases.  The cluster is elongated in the direction of a large
filament -- clearly visible at $z=0.5$ (Figure~2) -- along which sublcumps
are accreted onto the cluster. There are, however, noticeable differences
at both epochs in the substructures present in the various
simulations. These differences are due to discrepancies in the boundary
conditions (assumed to be isolated in Navarro, Steinmetz and Wadsley and
periodic in the rest), in the treatment of tidal forces, and in the
effective timing within the different simulations.

The models have been evolved for at least 21 expansion factors and
inaccuracies in the initial conditions, tidal forces, or integration errors
in the linear regime lead to a lack of synchrony at later times. These
timing discrepancies are manifest in the differing relative positions of
some subclumps at $z=0.5$ and are still apparent at $z=0$. For example,
there are two distinct substructures in Figures~1 and~8 to the NW of the
main clump at $z=0$ in slightly different positions in Bryan, Cen, Jenkins,
Owen, Pen and Warren. In Couchman, Evrard, Gnedin, Steinmetz and Yepes, one
of these substructures is already merging with the central clump, while in
Navarro both of them have merged. The differences in the overall shape and
orientation of the main concentration are largely due to a mismatch in the
epoch at which substructures are accreted.

With a uniform, 250 kpc, smoothing significant noise is visible in Owen.
Figure~8 shows the dark matter distribution at $z=0$, this time using the
smoothing considered as optimal by each simulator to display what they
considered to be real structure. There is a larger variety of structure in
these high resolution images than in the uniform smoothing case of
Figure~1. The simulations of Jenkins and Warren which have the largest
number of resolutions elements also have the largest number of satellite
structures. Varying numbers of these can be seen in other images, although
because of the slight timing differences they often appear in different
locations. The low and intermediate resolution simulations of Cen, Gnedin,
Owen and Yepes look quite similar when the standard smoothing or the
smoothing of choice is used.

\medskip
\noindent{\it Gas density.} 
\medskip

At the present epoch, the gas in all simulations (Figure~3) is rounder than
the dark matter -- a manifestation of the isotropic gas pressure -- and has
only a residual elongation along the accretion filament. Most of the
secondary clumps seen in the dark matter are also seen in the gas, but they
are clearly more diffuse. At $z=0.5$ (Figure~4), shortly before the final
large merger, the timing differences discussed above are quite apparent. In
some cases, the final major merger is already quite advanced but in others
two large subclumps are still clearly visible.

With this smoothing, the sampling in Owen is poor and Yepes' comparatively
low resolution is more apparent than in the corresponding dark matter
image. In Owen's case, the underlying asphericity of the SPH sampling is
lost when a fixed smoothing length is used; less noisy images result when
the geometry of the hydrodynamical sampling is maintained, as in the X-ray
image in Figure~7 below.

\medskip
\noindent{\it Gas temperature.} 
\medskip

The gas temperature images (Figures~5 and~6) show the most interesting
differences between the simulations. These are particularly striking at
$z=0.5$ when the slight timing differences apparent in the dark matter and
gas density plots produce quite dramatic differences in temperature
structure. In particular, in the simulations by Jenkins, Navarro and
Steinmetz, in which the final major merger has not yet occured at $z=0.5$,
an annulus of shock-heated material is evident between the two approaching
clumps, surrounding the axis of collision. In Evrard and Owen, the merger
is further advanced, but some residue of the annular structure remains. The
simulations of Bryan, Cen, Gnedin, and Yepes are yet further advanced and
while their temperature plots show similar departures from symmetry, the
annular structure is no longer evident. In Couchman and Wadsley, the merger
is nearly complete and the temperature distribution appears close to
spherically symmetric.

At $z=0$ the plots are broadly similar and most of the temperature
structure, both inside the cluster and in the outer regions of the
clusters, reproduces amongst the different simulations.

\medskip
\noindent{\it X-ray surface brightness.} 
\medskip

Like the dark matter distributions in Figure~8, the X-ray surface
brightness images in Figure~7 were generated using each simulator's
smoothing of choice. Since the X-ray surface brightness is calculated by
integrating ${\cal L}_X=\rho^2 T^{1/2}$, these images are similar to those
of the surface density (Figure~3), except that higher weight is given to
the central parts of the clumps. As a result, the central intensity is
strongly dependent on resolution.  In the region where the bulk of the
X-rays are produced, most images are quite similar. However, there are
large differences in the number and brightness of satellite structures.
Jenkins and Wadsley produced the largest number of such structures,
followed by Bryan and Steinmetz. Again, the substructures appear in
different places because of the timing discrepancies discussed above.

The high resolution grid codes of Bryan and Pen, and the intermediate
resolution grid code of Gnedin, show sharper structure in the low density
regions than the SPH codes, a reflection of their better treatment of
shocks in low density regions. Bryan's, Pen's, and all but Owen's SPH codes
have high central resolution and generally produce brighter central regions
than the low and intermediate resolution grid codes.

\subsection{Global properties} 

All simulators were requested to calculate global properties of their
clusters at the final time. The cluster was defined to be the material
lying within a spherical region around the center, of radius such that the
mean enclosed mass density is 201 times the critical value. In practice,
each simulator was left free to choose a preferred algorithm for locating
the cluster center since this choice is part of the uncertainties that we
are trying to assess. (The algorithms used in each case are described in
Section ~3.) The mean value of the cluster radius, averaged over all
simulations, was $r_{200}=2.70$ Mpc, with an rms scatter of 0.04.  We
display the global properties of the clusters in graphical form in
Figure~9, and discuss the different quantities one at a time. In each panel
in this figure, the simulations are arranged, from left to right, in order
of decreasing resolution, which is taken to be the maximum of the spatial
resolution for the gas at cluster center (column~3 of Table~1) and the
gravitational force resolution (column~6 of Table~1). Open circles are used
to represent SPH simulations and filled circles grid based simulations.

The simplest property of the cluster is its total mass, and all simulations
agreed on a value just over $10^{15}$M$_\odot$, to within better than
10\%. Differences arise from resolution and timing effects. Among the
grid-based codes, there is a clear trend of decreasing mass with decreasing
resolution. The timing discrepancies discussed in Section 4.1 affect the
position of substructures and thus the estimates of the cluster mass. For
example, among the SPH codes, Jenkins, Navarro, and Steinmetz find slightly
smaller masses than the others because of a significant lump clearly
visible in Figures~7 and~8 which falls just outside the cluster boundary in
their simulations but just inside it in the others. This is a result of the
late formation of their clusters, apparent in Figure~2.

The velocity dispersion of the dark matter particles within the cluster is
also reproduced to better than 10\% in all simulations except Yepes' whose
low value reflects the low force resolution of his simulation. (The
quantity plotted is the one-dimensional velocity dispersion, calculated as
$\sigma/\sqrt3$, where $\sigma$ is the full three-dimensional velocity
dispersion in the rest frame of the cluster.) There is some tendency for
the simulations which produced the largest total masses also to give the
largest velocity dispersions, but the correspondance is not exact,
reflecting the fact that the cluster is not in virial equilibrium to better
than 10\% and that the actual virial ratio depends on the detailed
positions of infalling clumps.

\hbox{~}
\vskip 2mm
\centerline{\psfig{file=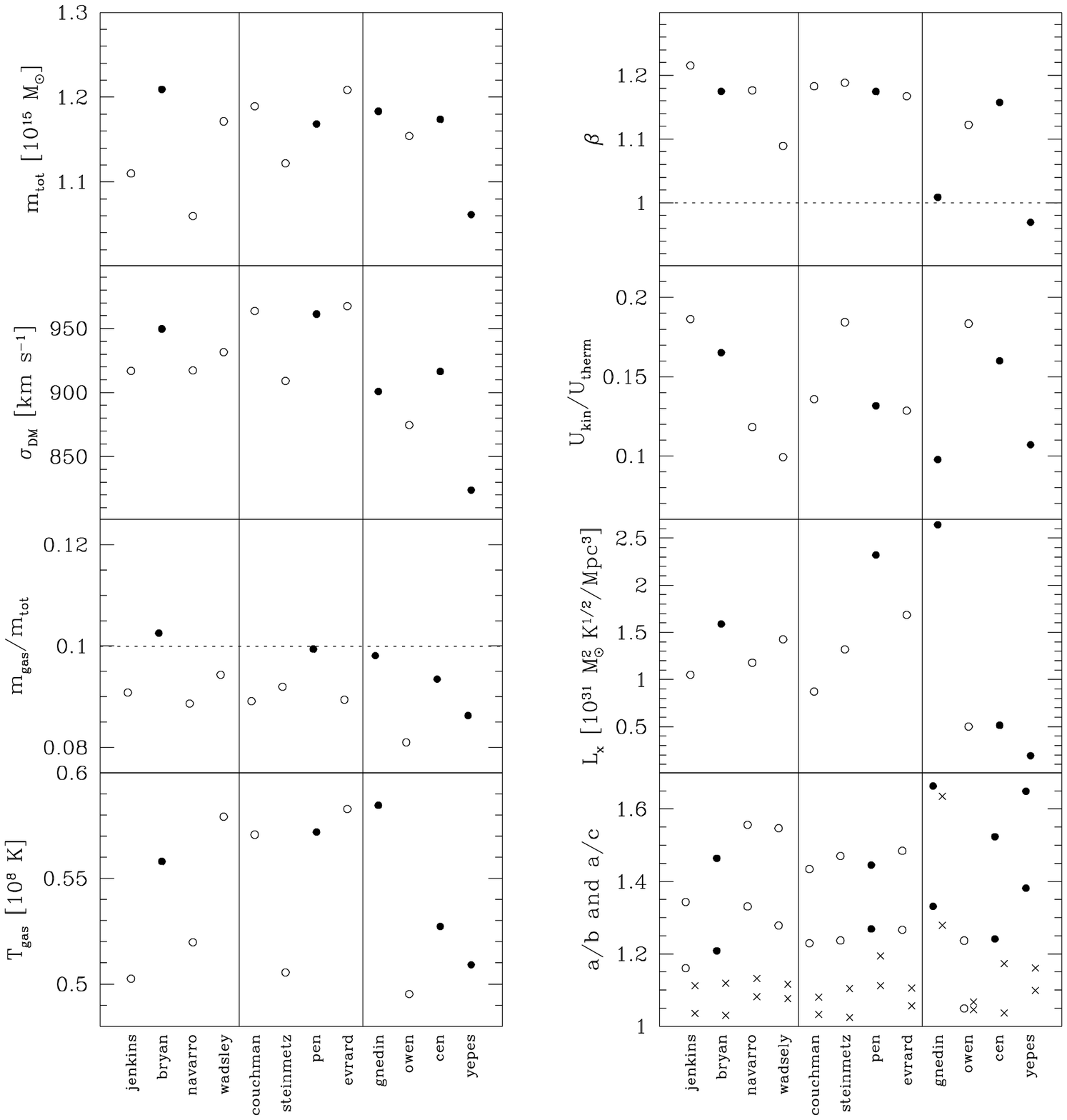,height=10.5cm}}
\noindent{
\scriptsize \addtolength{\baselineskip}{-3pt}
{\bf Fig.~9.} Global properties of the cluster in the various
simulations. All quantities are computed within the virial radius. From top
to bottom, the left column gives the values of: (a) the total cluster mass;
(b) the one-dimensional velocity dispersion of the dark matter; (c) the gas
mass fraction; (d) the mass-weighted gas temperature. Also from top to
bottom, the right column gives the values of: (a) $\beta=\mu
m_p\sigma_{DM}^2/3kT$; (b) the ratio of bulk kinetic to thermal energy in
the gas; (c) the X-ray luminosity; (d) the axial ratios of the dark matter
(circles) and gas (crosses) distributions. In each panel, the
models are arranged, from left to right, in order of decreasing resolution
which is taken to be the maximum of the spatial resolution for the gas at
cluster center (column~3 of Table~1) and the gravitational softening length
(column~6 of Table~1). Open circles are used to represent SPH simulations
and filled circles grid based simulations.
\vskip 2mm
\addtolength{\baselineskip}{3pt}
}

As might be expected, although still quite good, the agreement on the
properties of the gas is noticeably worse. The total amount of gas is most
interestingly expressed as a fraction of the total cluster mass. The
overall gas fraction in the model universe is 10\% and the highest
resolution grid simulations find a cluster gas fraction which is almost
exactly equal to this. Lower resolution grid simulations find progressively
smaller gas fractions, reflecting differential resolution effects in the
treatment of dark matter and gas in these codes. There is excellent overall
agreement among the SPH models except for the lowest resolution one:
everyone except Owen finds a gas fraction very close to 0.09. There seems
to be a systematic offset between the SPH models and the highest resolution
grid models but there is no clear indication of what may be causing this
difference. We explore this issue further in the next section by
considering the radial dependence of the gas fraction.

As a measure of temperature, all simulators calculated a mean,
mass-weighted temperature for all the gas within the cluster.  Everyone
found a value between 4.9 and $5.9\times 10^7$K.  Figures~5 and~6 suggest
that some of the differences result from the timing differences discussed
above which produce slightly different histories for the clusters just
prior to $z=0$. It is encouraging that the rms scatter in the measured mean
temperature is less than $\pm 7$\%. An even smaller scatter, $\pm 5\%$, is
found for the ratio of specific dark matter kinetic energy to gas thermal
energy, $\beta=\mu m_p\sigma_{DM}^2/3kT$, if Yepes' low value is
excluded. (Here $\mu$ denotes the mean molecular weight and $m_p$ the
proton mass.) This is further evidence that the differences in velocity
dispersion and temperature result from slightly different dynamical
histories rather than from differences in the treatment of the gas. Most
simulations give $\beta\simeq 1.17$, indicating that non-thermal or bulk
turbulent motions contribute to the support of the gas. The ratio of bulk
kinetic to thermal energy in the gas, plotted in the next panel of
Figure~9, is indeed about 15\%. This ratio correlates well with $\beta$,
but with small residuals that reflect slight departures from virial
equilibrium. The agreement on the values of $\beta$ and $U_{kin}/U_{therm}$
indicates that the shock capturing properties and the efficiency with which
infall kinetic energy is thermalized in shocks is similar in the SPH and
grid-based codes, at least in the regime explored in this simulation.  Most
simulators obtained values of about $3.6\times 10^{62}$ergs for the bulk
kinetic energy, except Gnedin and Wadsley who found values about 25\%
smaller. Lower turbulent energies probably result from some combination of
smaller noise-induced motion, greater viscous damping, and, in Wadsley's
case a more dynamically advanced state, but the actual factors responsible
in each case are unclear.

There is substantially less consensus about the estimated X-ray
luminosities of the cluster, calculated approximately as $\int
\rho^2T^{0.5}dV$, and so given in units of
M$_\odot^2$K$^{1/2}$Mpc$^{-3}$. The values found span a range of a factor
of $\sim 10$, or a factor of $\sim 5$, if we exclude Yepes' low resolution
model. Resolution effects also account for the small values obtained by Cen
and Owen.  The largest values were obtained in the intermediate and high
resolution grid-based simulations of Gnedin and Pen respectively, and in
Evrard's SPH simulation. Evrard's and Wadsley's models produced larger
X-ray luminosities than other SPH simulations because their clusters are in
a slightly more advanced (and more active) dynamical state. The higher
luminosities from the high resolution grid-based codes are due to their
slightly more concentrated central gas distributions (see Figure~3 and
Section~4.3 below). Because the total X-ray luminosity is sensitive to the
structure of the inner few hundred kiloparsecs of the cluster, it
fluctuates quite strongly in time and is very sensitive to simulation
technique.

As a final test, we compare the shapes of the clusters, as measured by the
inertia tensors, ${\cal I} = \sum m_{i}{\bf x}_i{\bf x}_i/\sum m_i$, for
both the dark matter and the gas within the spherical region which defines
the cluster. We label the eigenvalues of this tensor $a^2 > b^2 > c^2$, and
define the axial ratios to be $a/b$ and $a/c$. As can be seen in Figures~1
and~3, the cluster is aspherical, with the orientation of its longest axis
reflecting its formation by infall along a filament. The axial ratios shown
in Figure~9 show, in fact, that the cluster is triaxial. The dark matter
distribution (circles in Figure~9) is considerably more aspherical than the
gas (crosses) in all cases except Gnedin's. There is generally good
agreement amongst the different simulations, although there are a few
anomalies. For example, Owen finds the smallest axial ratios for the dark
matter distribution even though the inner regions of his cluster appear
quite elongated in Figure~1. This is probably because of the relatively
large contribution to ${\cal I}$ from the largest infalling clump which
lies close to $r_{200}$ in his simulation. Gnedin's dark matter and gas
distributions are considerably more aspherical than the rest.

\subsection{Radial profiles}

In order to perform more detailed quantitative comparisons of cluster
structure, each simulator was asked to provide the radial profiles of a
number of cluster properties. These were averaged in a specified set of
spherical shells centered at the position deemed by each simulator to be
the cluster center (cf \S2.2).  Simulators were also asked to specify the
effective resolution of their simulation, and throughout this section we
plot data for each model only at radii larger than this. Most simulators
identified the resolution of their simulation with the effective
gravitational force resolution (see Table~1), but the following specified
different values: Cen (200 kpc), Owen (500 kpc), Pen (50 kpc), Steinmetz
(35 kpc), and Yepes (400 kpc).  In the figures that follow, the data points
are slightly displaced from the bin centers for clarity, and we use a solid
line to show the mean profile obtained by averaging all the data plotted in
each bin. At the top of each diagram we plot the residuals from this mean,
defined in most cases as ${\rm ln} x - <{\rm ln} x>$, where $x$ is the
property of interest and the brackets denote the average of the data points
in each bin. This definition applies to the residuals of all the radial
profiles, except those of the normalized baryon fraction (Figure~13), which
we define as $(x -<x>)/<x>$, and those of the radial velocity profiles
(Figures~14 and~15) and the entropy (Figure~18) which we define as
$x-<x>$. The sampling errors in the estimates of the different cluster
properties are largest in the innermost bin plotted where they are
typically less than about 10\%.

\hbox{~}
\vskip 2mm
\centerline{\psfig{file=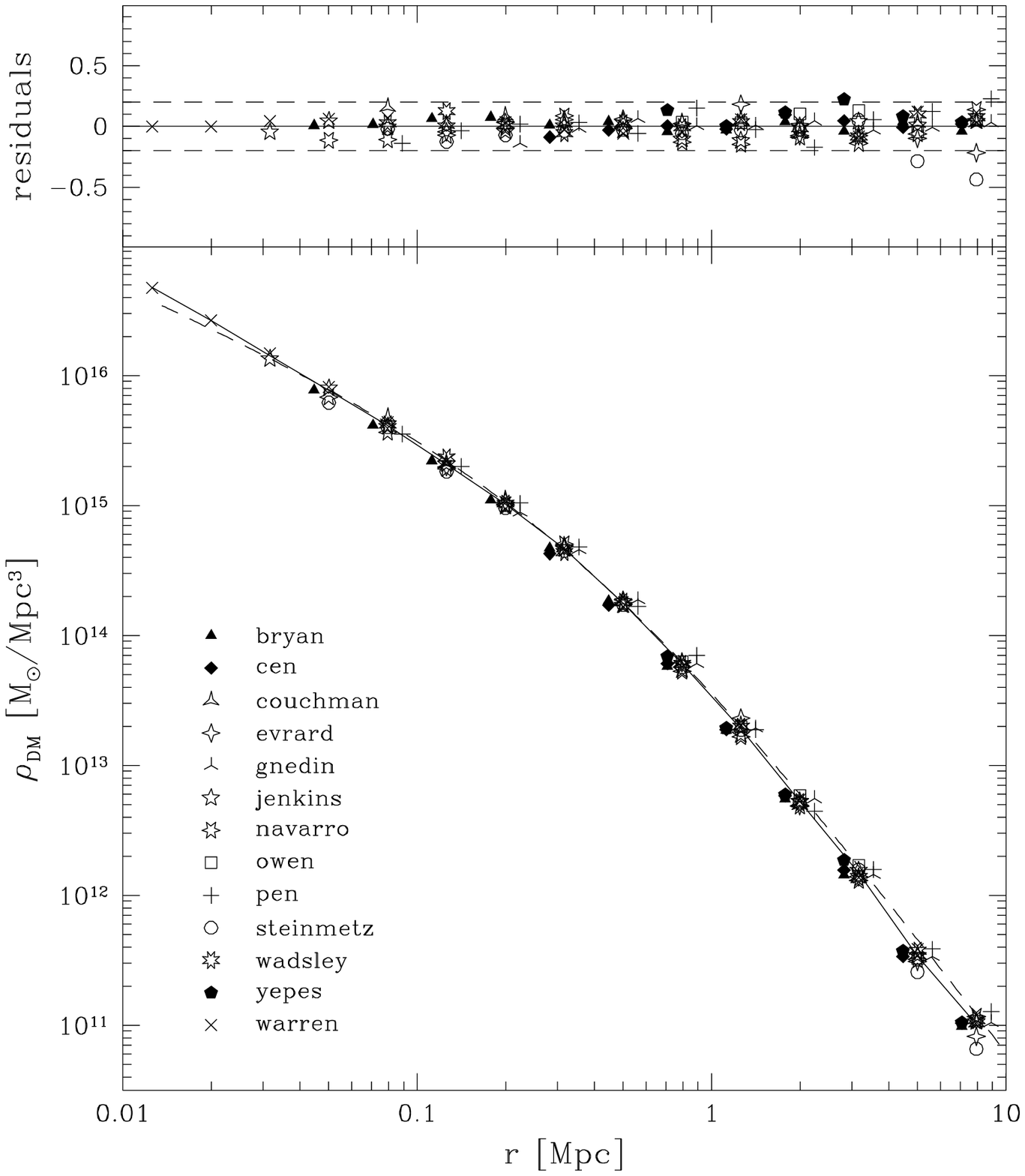,height=11.0cm}}
\noindent{
\scriptsize \addtolength{\baselineskip}{-3pt}
{\bf Fig.~10.} Projected dark matter density at $z=0$. The images, covering the inner 8
Mpc of each simulation cube, have been smoothed using the standard Gaussian
filter of 250 kpc half-width described in the text. Wadsley's simulation,
not shown here or in Figure~2, has a similar appearance to Couchman's.
\vskip 2mm
\addtolength{\baselineskip}{3pt}
}

We begin by comparing the dark matter density profiles plotted in
Figure~10. In general there is very good agreement between the different
calculations over the regions resolved by each simulation.  The two highest
resolution models, Warren's pure N-body simulation, and Jenkins'
AP$^3$M/SPH simulation, agree extremely well at all radii.  The residuals
plot shows that all simulations agree to within $\pm 20\%$ at all radii.
The dark matter profile in this cluster is well fit by the analytic form
proposed by Navarro, Frenk and White (1995), all the way from 10 kpc to 10
Mpc. This fit is shown as a dashed line in Figure~10 and corresponds to a
value of the concentration parameter, $c=7.5$, appropriate to a typical
isolated halo of this mass in an $\Omega=1$ CDM model (Navarro, Frenk \&
White 1997).  Only in the very center is there a slight indication that the
true profile, as defined by Warren's model, might be steeper than the
analytic form.

The dark matter velocity dispersion profiles in Figure~11 confirm that all
the codes give very similar results for the dynamical properties of the
dark matter. (Note the very different dynamic ranges in Figures~10 and
~11.) Except for the last bin, which is particularly affected by noise
arising from subclustering, the scatter in the velocity profiles is
comparable to that in the mass profiles, about 20\%.  The velocity
dispersion profile rises near the center, has a broad peak around 100-500
kpc, and declines in the outer parts. Warren's N-body model and all the
high resolution SPH and grid models resolve the inner rising part of the
velocity dispersion profile.

\hbox{~}
\vskip 2mm
\centerline{\psfig{file=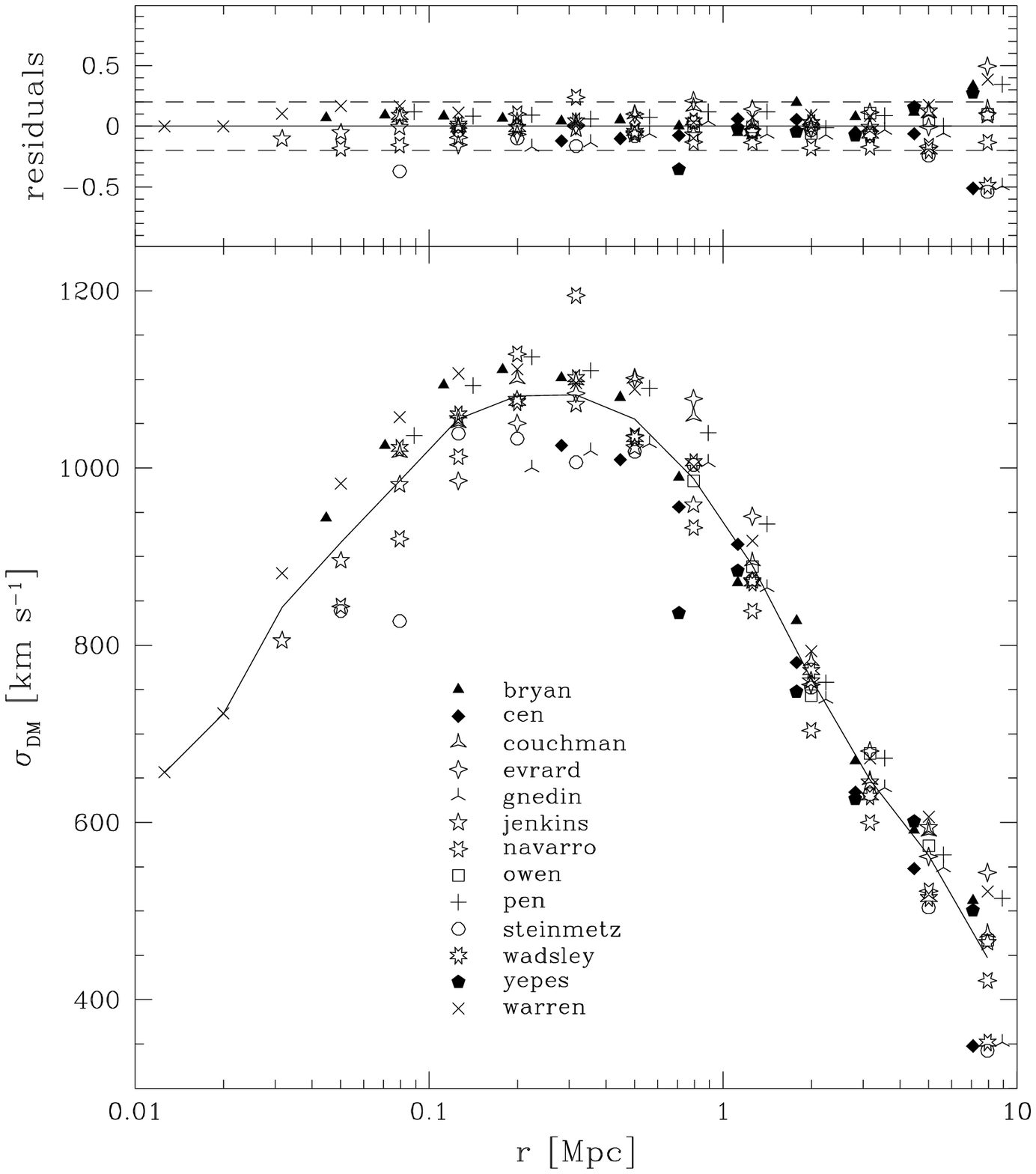,height=11.0cm}}
\noindent{
\scriptsize \addtolength{\baselineskip}{-3pt}
{\bf Fig.~11.} The dark matter velocity dispersion profile. The
quantity shown in the main plot is the one-dimensional velocity dispersion,
calculated as $\sigma/\sqrt3$, where $\sigma$ is the full three-dimensional
velocity dispersion. See the caption to Figure~10 for a description of the
symbols and other details of the plot.
\vskip 2mm
\addtolength{\baselineskip}{3pt}
}

The agreement of the gas density profiles (Figure~12) is less good but is
still quite impressive. In the fixed Eulerian grid models of Cen and Yepes,
the gas density at their innermost point is somewhat low, whereas the
corresponding dark matter matter densities agree well with the other
calculations. This shows, unsurprisingly, that the resolution of such codes
is somewhat poorer for the hydrodynamics than for the N-body dynamics.
Bryan's multilevel grid code produces results that agree quite well with
other high resolution models, except that his two innermost points lie
slightly below those of the highest resolution SPH models. Gnedin's and
Pen's variable resolution grid codes give higher than average gas densities
in the 200 to 600 kpc range and, as a result, their density profiles are
steeper than the others. Pen's simulation produced a more pronounced core
structure within about 150 kpc than all other models while Gnedin's
simulation shows the largest departures from the mean profile. There is
also significant scatter among the results of the various SPH codes, Evrard
finding systematically high gas densities in the inner cluster and Couchman
finding systematically low values. It seems likely that at least some of
these differences result from the differences in the timing of cluster
collapse noted in Section 4.1, rather than from differences in the
treatment of the hydrodynamics between the different techniques and
implementations although the two cannot be clearly disentagled. For
example, Couchman's gas densities between 80 and 200 kpc are somewhat lower
than those obtained by Jenkins using the same SPH implementation but
different numbers of particles and integration parameters. The dashed line
in Figure~12 is the mean dark matter density profile reproduced from
Figure~10. It is interesting that this profile is substantially steeper
than the mean gas density profile in the inner cluster, indicating that the
gas has developed a much more clearly defined ``core'' than the dark
matter.
\hbox{~}
\vskip 2mm
\centerline{\psfig{file=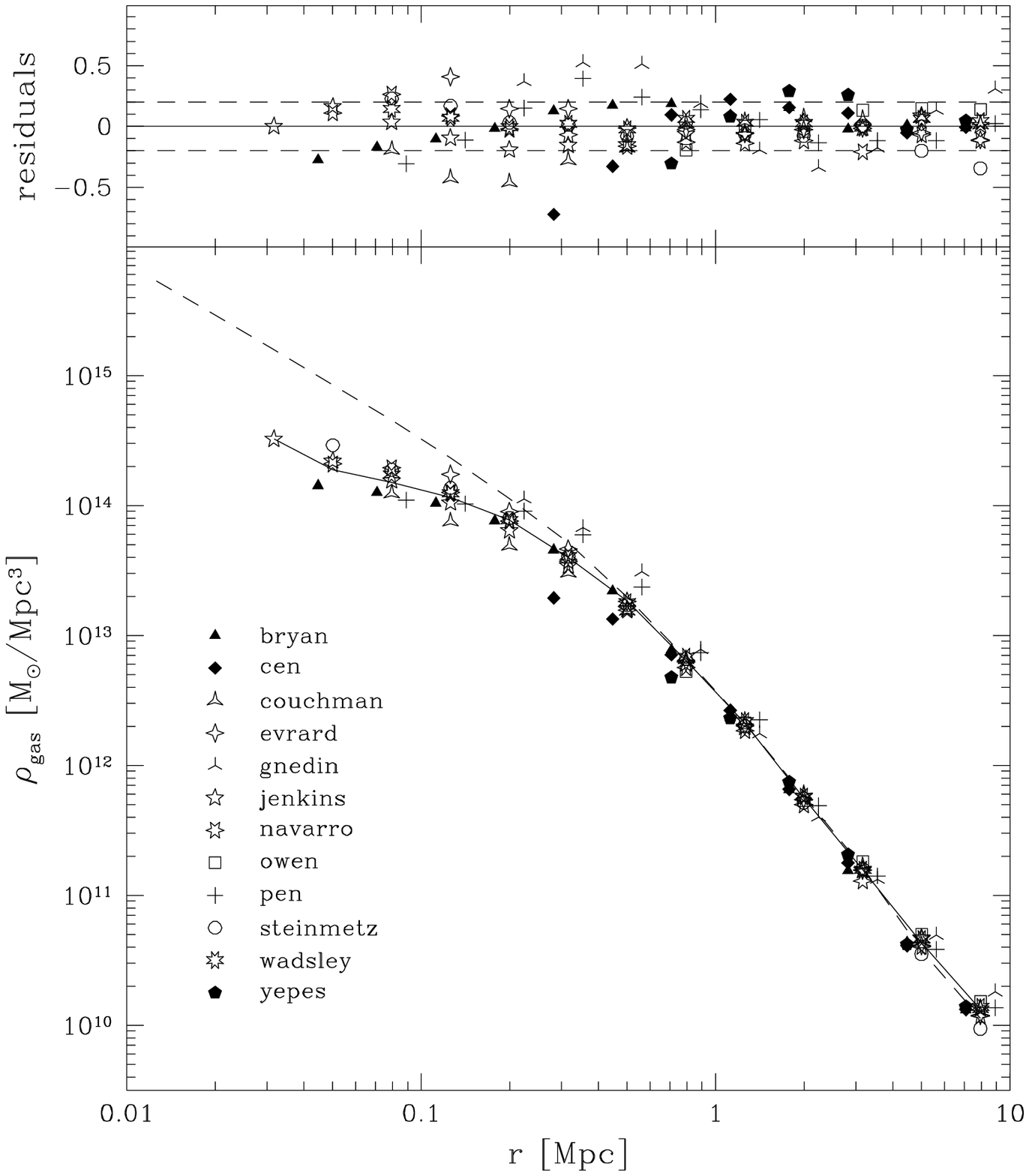,height=11.0cm}}
\noindent{
\scriptsize \addtolength{\baselineskip}{-3pt}
{\bf Fig.~12.} The gas density profile. The dashed line shows 
the mean dark matter density profile reproduced from Figure~10. See the
caption to Figure~10 for further details.
\vskip 2mm
\addtolength{\baselineskip}{3pt}
}

The discrepancies between the various dark matter and gas density profiles
become clearer when we examine the variation of the normalized gas
fraction, $\Upsilon=M_{gas}(<r)/(\Omega_b M_{tot}(<r))$, with radius
(Figure~13). There is considerable scatter ($\sim 50\%$) in this gas
fraction over the 100~kpc to 1~Mpc range, Gnedin finding the highest values
and Couchman, Cen, and Yepes the lowest. Well inside the virial radius,
three of the grid models, Bryan's, Pen's, and especially Gnedin's, rise
above $\Upsilon =1$, the mean for the simulation as a whole. At larger
radii, Pen's values fall slightly below this mean, while Bryan's and
Gnedin's remain slightly above unity well beyond the virial radius. By
contrast, none of the SPH models ever rise above $\Upsilon =1$ and they all
give very similar results at the virial radius, similar also to the
Eulerian models of Cen and Yepes. Although relatively small, these
discrepancies appear to reflect a systematic difference in the final
distribution of the gas between the three intermediate and high resolution
grid simulations (Gnedin, Pen and Bryan) and the rest of the models.
Particularly puzzling is the excursion towards large values of $\Upsilon$
seen by Gnedin at $r\simeq 1$ Mpc and the fact that Bryan and Gnedin obtain
values of $\Upsilon > 1$ beyond three virial radii where one might expect
the mixture of dark matter and gas to attain the mean universal
value. Note, however, that the deviations from $\Upsilon=1$ at large radii
are quite small.

The infall patterns of dark matter and gas around the cluster (i.e. the run
of peculiar radial velocities) are illustrated in Figures~14 and~15. For
the dark matter, the radial velocity profiles are quite similar: net infall
is seen in most models beyond $\sim 700$ kpc, except in Pen's case, in
which the radial velocity remains close to zero until about twice this
radius. There are larger differences in the radial velocity profiles for
the gas. In most models the gas is infalling over the same range of radii
as the dark matter, but at a slightly lower speed. Gnedin's model is
anomalous in this respect. In Pen's simulation, on the other hand, there is
a small net outflow of gas at $\sim 1$ Mpc.  The scatter in the radial
velocity profiles beyond $\sim 1$ Mpc is about 200 \kms.

\hbox{~}
\vskip 2mm
\centerline{\psfig{file=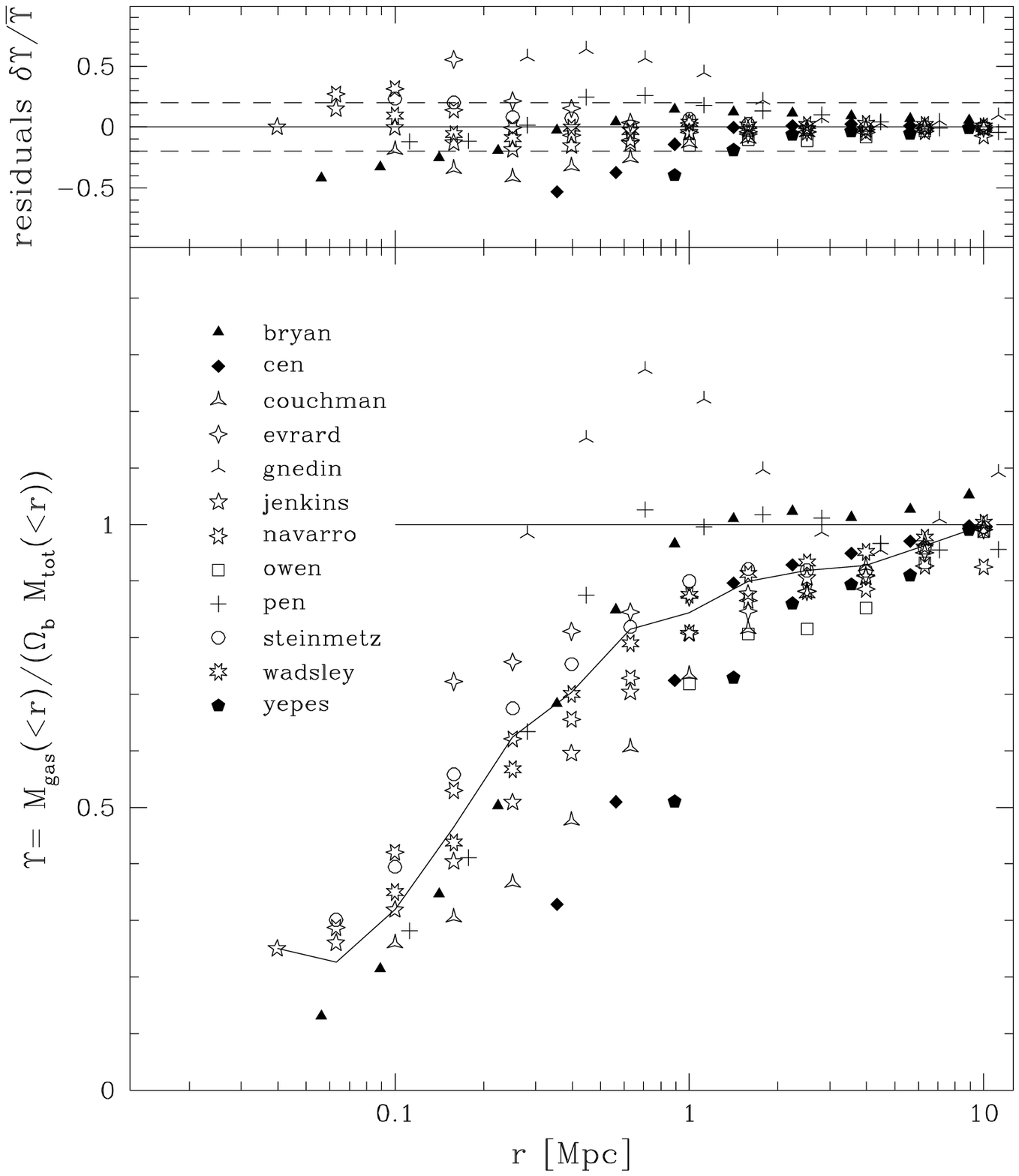,height=11.0cm}}
\noindent{
\scriptsize \addtolength{\baselineskip}{-3pt}
{\bf Fig.~13.} The radial dependence of the gas fraction. 
The quantity plotted is the gas fraction normalized to the value for
the simulation as a whole (10\%). See the caption to Figure~10 for further 
details, but note that in this Figure the residuals are defined as
$(\Upsilon -<\Upsilon>)/<\Upsilon>$.
\vskip 2mm
\addtolength{\baselineskip}{3pt}
}

\hbox{~}
\vskip 2mm
\centerline{\psfig{file=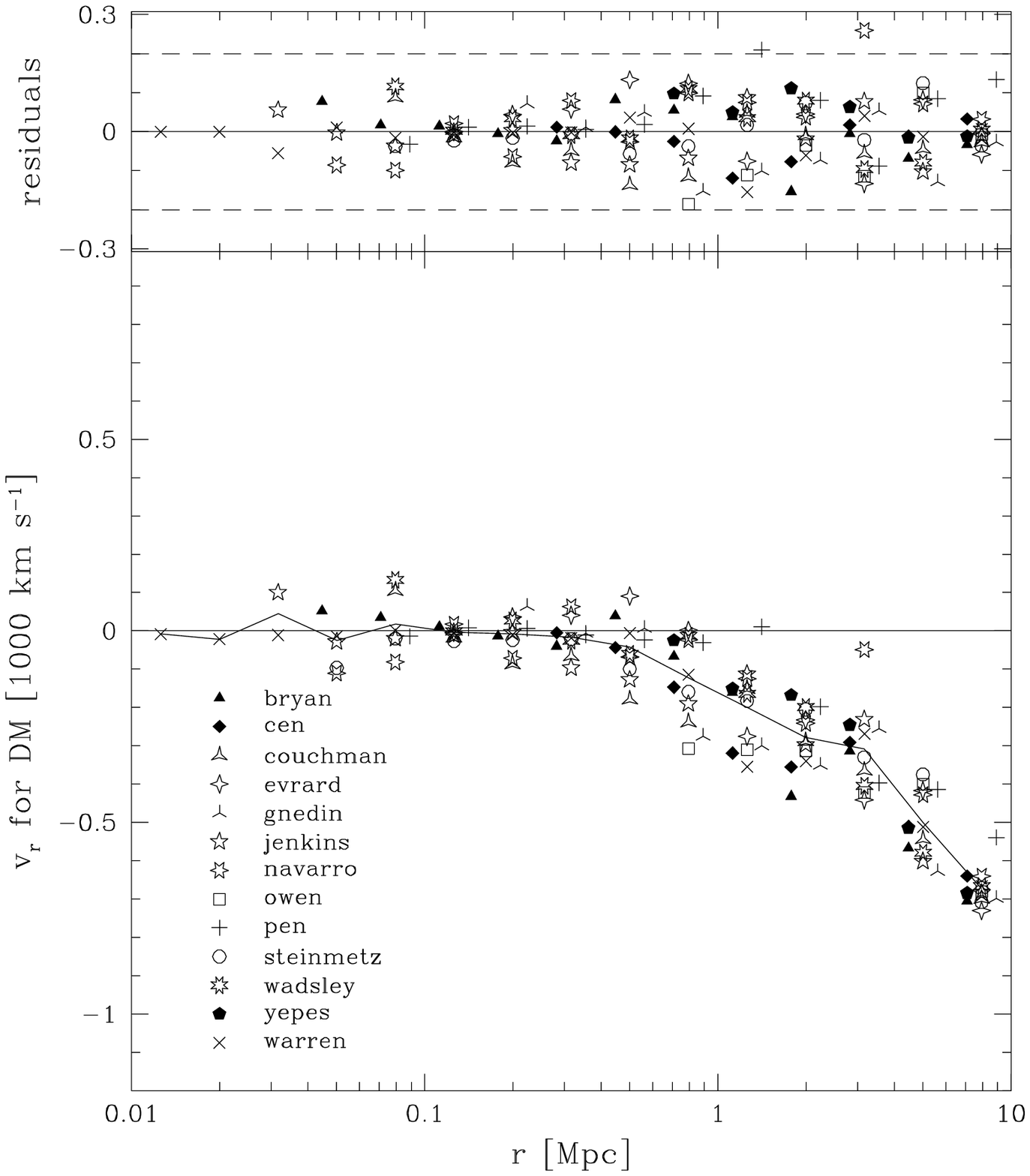,height=11.0cm}}
\noindent{
\scriptsize \addtolength{\baselineskip}{-3pt}
{\bf Fig.~14.} The radial velocity profile of the dark matter. 
Velocities are computed in the rest frame of the cluster and do not include
the Hubble expansion. See the caption to Figure~10 for further
details, but note that in this Figure the residuals are defined as $v_r
- <v_r>$.
\vskip 2mm
\addtolength{\baselineskip}{3pt}
}

\hbox{~}
\vskip 2mm
\centerline{\psfig{file=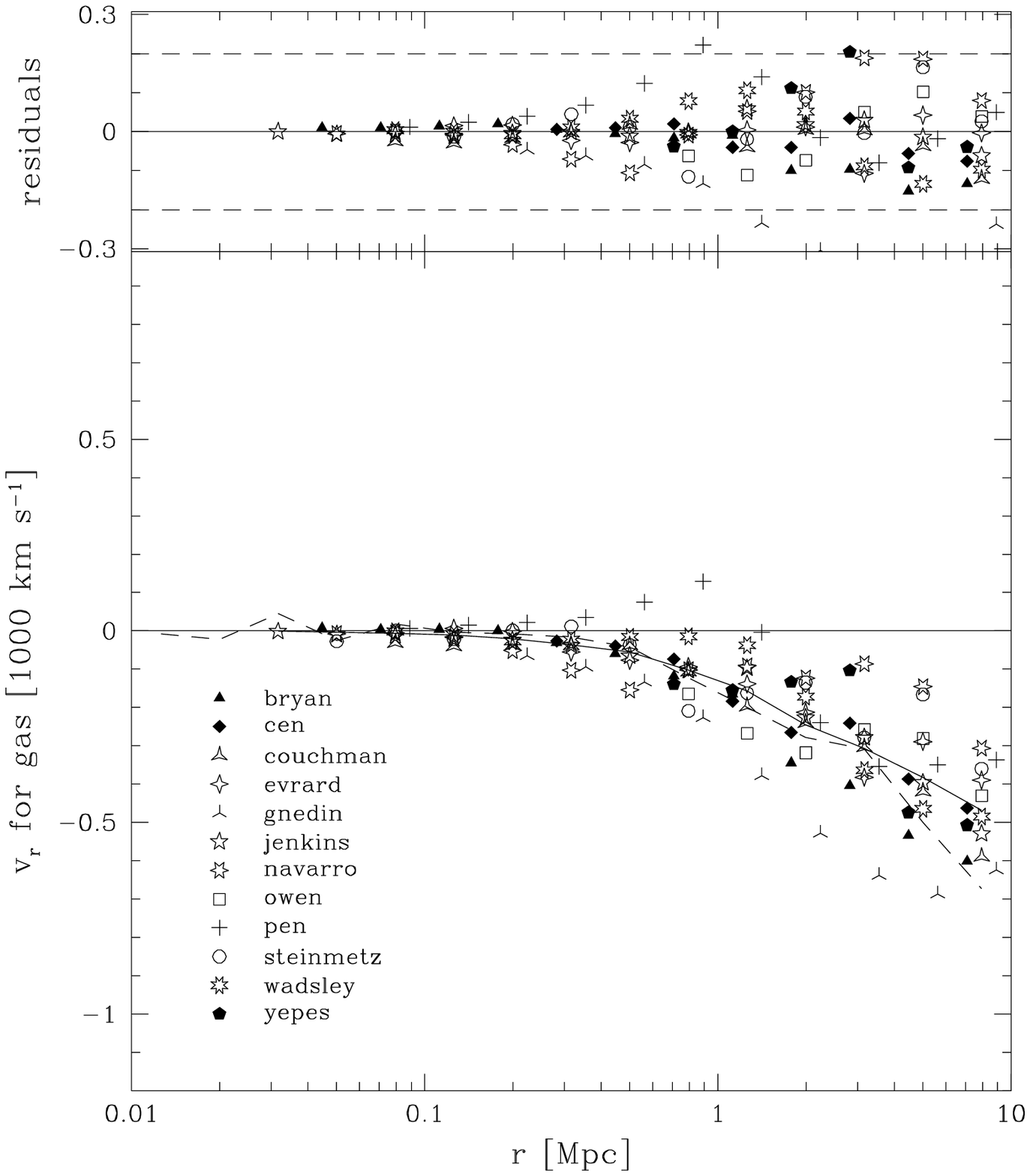,height=11.0cm}}
\noindent{
\scriptsize \addtolength{\baselineskip}{-3pt}
{\bf Fig.~15.}The radial velocity profile of the gas. 
Velocities are computed in the rest frame of the cluster and do not include
the Hubble expansion. The dashed line is the dark matter radial velocity
profile from Figure~14. See the caption to Figure~10 for further
details, but note that in this Figure the residuals are defined as $v_r 
-<v_r>$. 
\vskip 2mm
\addtolength{\baselineskip}{3pt}
}

The differences between the codes become most obvious if we look at various
thermodynamic properties of the gas. Figures~16 through~19 show radial
profiles for the pressure, temperature, entropy and X-ray emissivity. (Of
these, only the temperature was calculated directly by the simulators; the
other quantities were derived from the binned values of temperature,
density, and X-ray luminosity.)  Agreement among the pressure profiles is
reasonably good. This reflects the good agreement of the dark matter
density profiles and so of the gravitational potential wells, together with
the fact that the intracluster gas is close to hydrostatic equilibrium in
all cases. In the central regions, simulations which give flatter gas
density profiles (e.g. Bryan, Pen, Couchman, and Cen) can be clearly seen
to give correspondingly higher temperatures, as required in order to
maintain comparable pressure gradients. The differences in the temperature
profiles appear substantially larger than in other quantities, but this is
in part a reflection of the smaller dynamic range of the plot. There is a
suggestion that the temperature structure in the inner parts may be
systematically different in the SPH and grid models. In the former, the
temperature profile is flat or slowly declining towards the inner regions,
but in the grid simulations, the temperature is still rising at the
innermost point plotted, a trend that is particularly noticeable in Bryan's
simulation.  In the outer cluster, the temperature profile drops
substantially in all cases.
\hbox{~}
\vskip 2mm
\centerline{\psfig{file=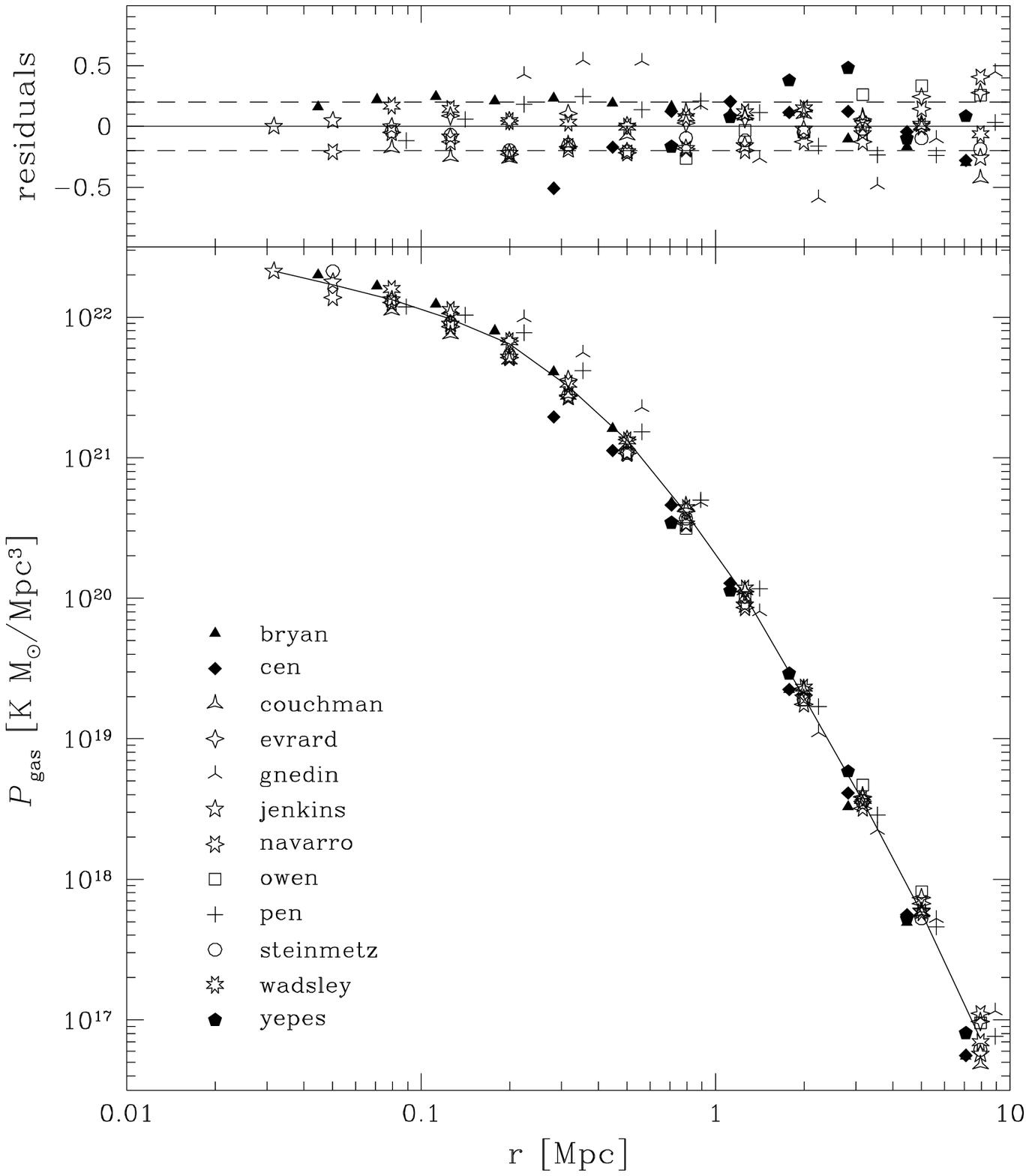,height=11.0cm}}
\noindent{
\scriptsize \addtolength{\baselineskip}{-3pt}
{\bf Fig.~16.} The gas pressure profile. See the caption to
Figure~10 for further details.
\vskip 2mm
\addtolength{\baselineskip}{3pt}
}

\hbox{~}
\vskip 2mm
\centerline{\psfig{file=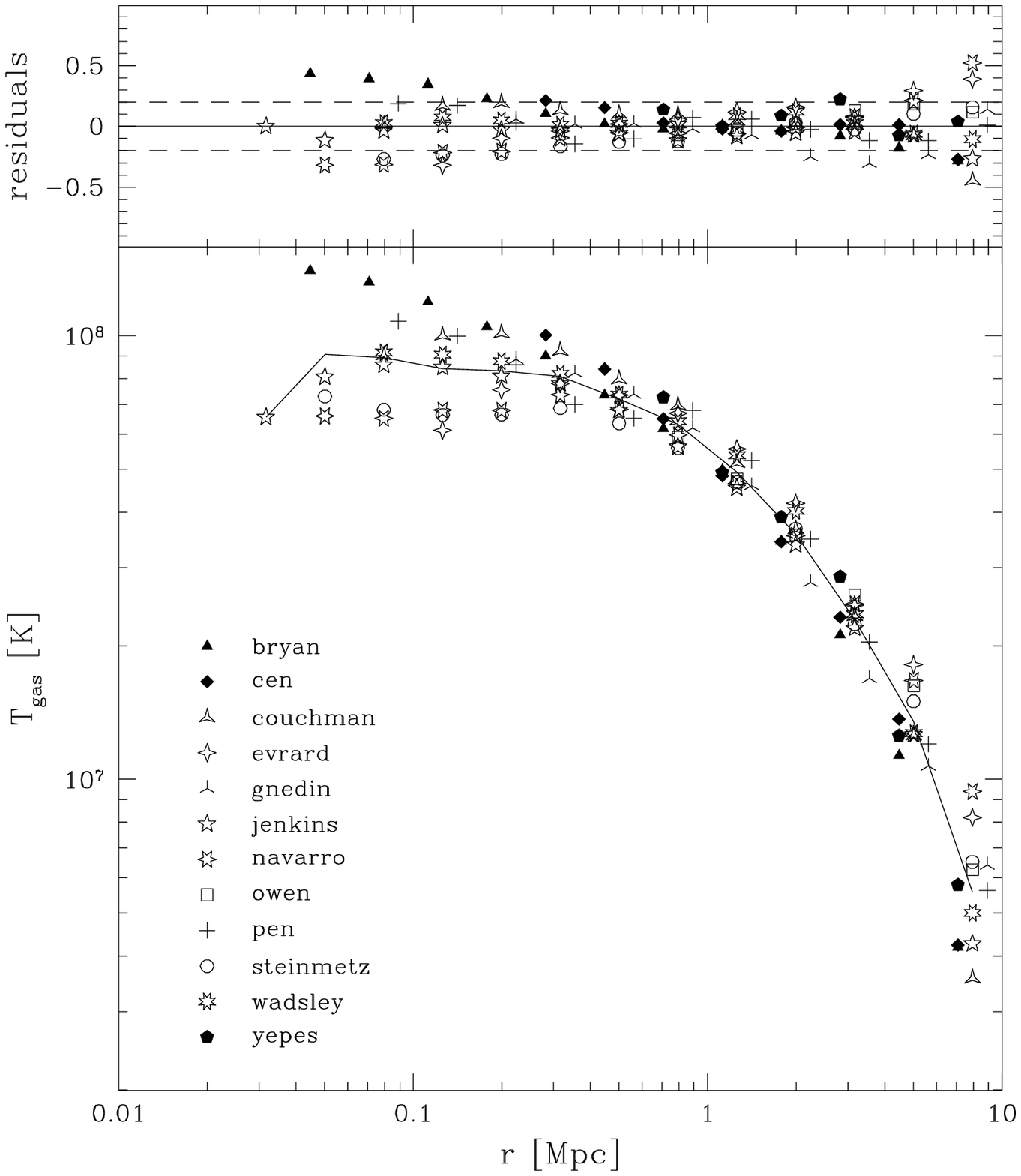,height=11.0cm}}
\noindent{
\scriptsize \addtolength{\baselineskip}{-3pt}
{\bf Fig.~17.}The gas temperature profile. The quantity plotted
is the mass-weighted temperature. See the caption to Figure~10 for further 
details. 
\vskip 2mm
\addtolength{\baselineskip}{3pt}
}

The entropy and X-ray luminosities show the patterns expected for
quantities derived in a simple way from the density and temperature of the
gas. Note, in particular, that the entropy (defined as $s={\rm
ln}(T/\rho_{gas}^{2/3})$ decreases systematically towards the center in all
the SPH models, but that this decline is less obvious in the grid models
and is, in fact, absent in the central parts of Bryan's simulation in which
the entropy remains approximately constant within $\sim 200$ kpc.  This
difference might reflect differences in the way in which shocks are treated
in the SPH and grid codes; however, the effect is small and occurs at the
resolution limit of the grid simulations. Finally, Figure~19 shows the
contribution to the total X-ray luminosity per logarithmic interval in
radius, $4\pi r^3 {\cal L}_X$. Most of the X-ray luminosity is produced in
the radial range 200-500 kpc. This region was well resolved by all the SPH
simulations and by Bryan's model which agrees remarkably well with them. On
the other hand, Cen, Owen and Yepes did not resolve this radial range and,
as a result, their total luminosities ended up being smaller than
average. The large gas densities found by Gnedin and Pen in this range
account for their larger than average total luminosities (see Figure~9).
Thus, the large scatter in total X-ray luminosity seen in Figure~9 results
partly from the low resolution of some of the models and partly from the
unusually high gas densities found by the two deformable grid
models. Between $\sim 1$ Mpc and the virial radius the different models
agree better although the scatter in $4\pi r^3 {\cal L}_X$ is still larger
than in all other properties.
\hbox{~}
\vskip 2mm
\centerline{\psfig{file=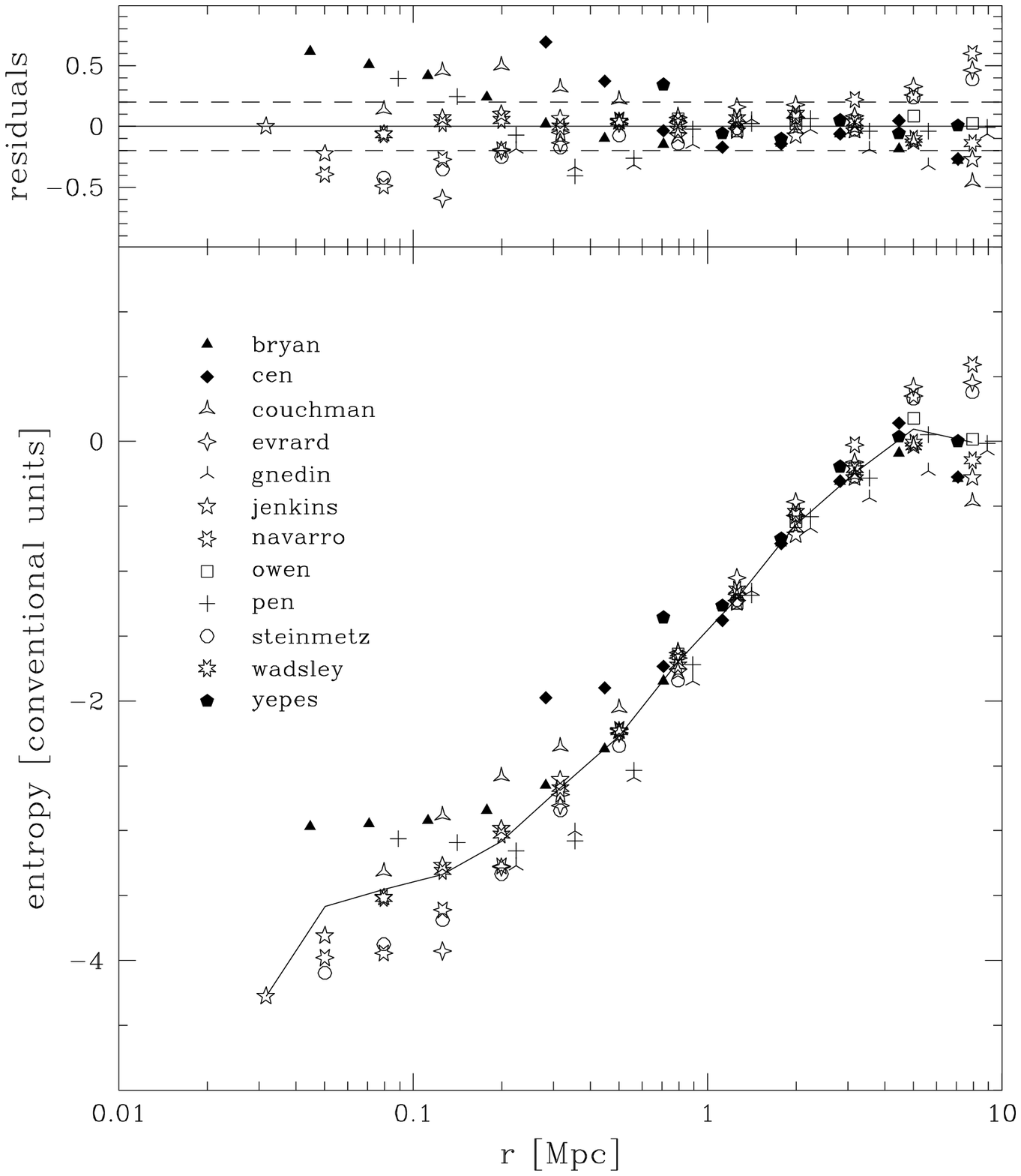,height=11.0cm}}
\noindent{
\scriptsize \addtolength{\baselineskip}{-3pt}
{\bf Fig.~18.}The radial variation of the gas entropy. The
entropy is defined as $s={\rm ln}(T/\rho_{gas}^{2/3})$. See the caption to
Figure~10 for further details. 
\vskip 2mm
\addtolength{\baselineskip}{3pt}
}

\hbox{~}
\vskip 2mm
\centerline{\psfig{file=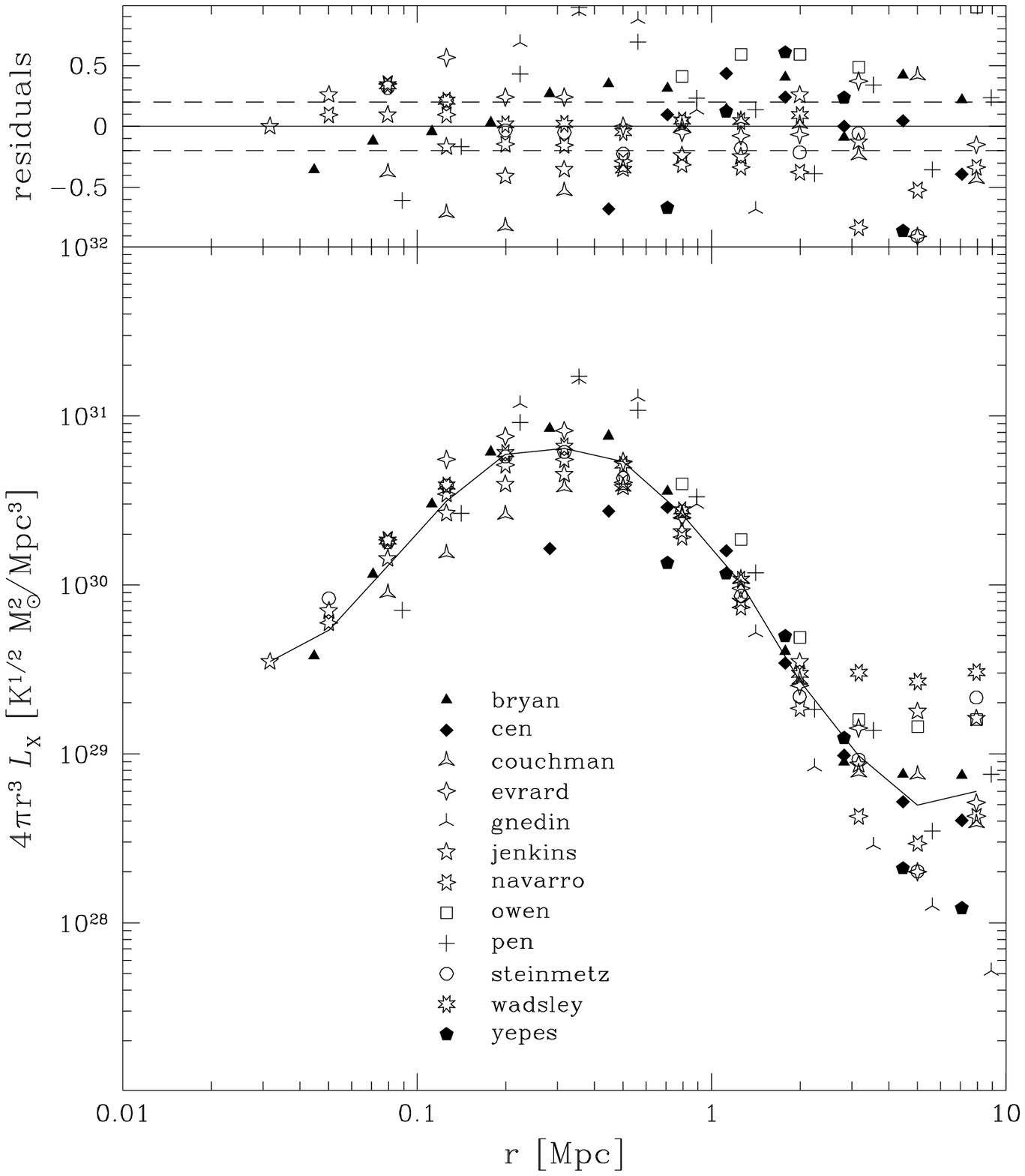,height=11.0cm}}
\noindent{
\scriptsize \addtolength{\baselineskip}{-3pt}
{\bf Fig.~19.} The X-ray luminosity profile. The quantity plotted
is $4\pi r^3 {\cal L}_X$, where ${\cal L}_X$ denotes the X-ray luminosity 
density in each bin. See the caption to Figure~10 for further details.
\vskip 2mm
\addtolength{\baselineskip}{3pt}
}

\section{Discussion and conclusions}

We have simulated the formation of an X-ray cluster in a cold dark matter
universe using 12 different cosmological gas dynamics codes that span the
range of numerical techniques and implementations presently in use. This
comparison aims to assess the reliability of current cosmological
simulations in the regime relevant to the bulk of the gas in galaxy
clusters, and to set a standard against which future techniques may be
tested. (The initial conditions and a selection of results are available at
http://star-www.dur.ac.uk/~csf/clusdata/ or by request from CSF.)  Because
our goal is to compare results in a realistic situation, only the initial
conditions were specified.  Other variables such as resolution, the
treatment of boundary conditions, and integration parameters were left to
the discretion of simulators. Our comparison therefore encompasses not only
the current variety of hydrodynamics techniques, but also the different
choices commonly made by individual authors for these variables. Seven of
the codes used for this comparison are based on the SPH technique, while
the other five are based on grid methods, employing either a fixed, a
deformable or a multilevel mesh.  The resolution of the simulations varied
from a few tens to a few hundreds of kiloparsecs. Although there is, of
course, no guarantee that any of the calculations gives the correct
solution to the problem, the agreement among the various simulations was
better than a pessimist might have predicted.  Nevertheless, some important
differences do exist.

The simulated cluster was chosen to have a mass comparable to the Coma
cluster at the present day and to appear fairly relaxed on visual
inspection. We assumed cold dark matter initial conditions, $\Omega=1$,
$h=0.5$, and a global baryon fraction of 10\%. The cluster formed by the
merging of subclumps infalling along a filament and experienced a final
major merger between $z=0.5$ and $z=0$. The final cluster mass, virial
radius, gas fraction within the virial radius, and mass-weighted gas
temperature, averaged over all the simulations, and the standard deviation
in each quantity are $M=1.1
\times 10^{15}$\Mo, $\sigma_M= 0.05\times 10^{15}$\Mo; $r_v=2.70$ Mpc,
$\sigma_{rv}= 0.04$ Mpc; $f_b=0.092$, $\sigma_{f_b}=0.006$; and $T=5.4
\times 10^7$K, $\sigma_T=0.34
\times 10^7$K respectively.

The properties of the cluster dark matter are gratifyingly similar in all
the models. The total mass and velocity dispersion agree to better than
5\%. The dark matter density and velocity dispersion profiles are also
similar and match the result of a higher resolution dark matter only
simulation. Over the regions adequately resolved in each simulation, the
scatter in these quantities, relative to the mean profile, is less than
about $\pm 20\%$ per logarithmic bin in radius. For the most part, this
scatter seems to be due to a slight asynchrony in the evolutionary state of
the models introduced by inaccuracies in the initial conditions, the
treatment of boundary conditions and of tidal forces and the integration
procedure. Thus, small subclumps often appear at different positions and
the timing of the final major merger differs slightly in the different
models.

There is less agreement on the gas properties of the cluster, although in
most cases they are quite similar. For example, all models agree to 10\%
(rms) on the gas mass and baryon fraction within the virial radius. In all
the SPH and all but one of the grid models, the gas is slightly more
extended than the dark matter. The scatter relative to the mean in
logarithmic radial bins seldom exceeds 20\% in the case of the density
profile and 30\% in the case of the gas mass fraction. There is no obvious
systematic difference in the gas density profiles produced by the SPH and
fixed grid models, but the deformable grid models produced somewhat larger
core radii. At large distances from the cluster, some of the grid based
models rise slightly above the theoretical expectation of a universal
baryon fraction.

The mean (mass-weighted) gas temperature is reproduced to within 6\% (rms)
by all the codes. The ratio of specific dark matter kinetic energy to gas
thermal energy is reproduced to a similar accuracy. The scatter per
logarithmic bin in the temperature profiles falls in the $\pm 20\%$ band
and is only slightly larger for the pressure profile. In the central
regions ($r\leq 100$kpc), however, the SPH codes produce a flat or slightly
declining temperature profile while all the grid codes produce a
temperature profile that is still rising at the resolution limit. The
entropy of the gas declines continuously from the virial radius to the
resolution limit but there is a suggestion that the entropy in the grid
codes may bottom out at small radii while it continues to decrease in the
SPH codes.

Amongst all the properties we have examined, the largest discrepancies
occur in the predicted cluster X-ray luminosity. This is proportional to
the square of the gas density and so is strongly dependent on resolution
and is further affected by variations in the potential produced by the
exact timing of the final major merger. The large range of effective
resolution in the various models gives rise to a factor of 10 variation in
the total X-ray luminosity. The luminosity per logarithmic interval in
radius peaks just outside the gas core radius. The eight simulations that
resolved this region (all the SPH and two of the grid models) show a much
narrower spread in total X-ray luminosity, amounting to a factor of 2.6 (or
1.8 if the most extreme model is excluded.)

We conclude that the different approaches to modelling shocks and other
hydrodynamical processes implicit in the diverse techniques employed in
this comparison give, in most cases, fairly consistent results for the
dynamical and thermodynamical properties of X-ray clusters. Variations
introduced by differences in the internal timing of the simulations tend to
be at least as important as variations in the treatment of
hydrodynamics. An illustration of this is the comparison of the simulations
of Couchman and Jenkins who used serial and parallel versions of
essentially the same code, but with different numbers of particles and
other simulation parameters. The final temperatures of their clusters
differed by 15\%, the X-ray luminosities by 20\%, and the bulk gas kinetic
energy by 15\%.

The conclusions discussed in this paper apply exclusively to the particular
case of a non-radiative gas. They cannot be extrapolated to other regimes
such as that appropriate to galaxy formation where gas cooling is a
dominant process. The behavior of the gas in this situation is determined
by the resolution of the simulation because the cooling rate depends
strongly on gas density. In addition, simulations of galaxy formation often
include algorithms to convert cold gas into stars and to model the feedback
processes associated with star formation. A comparison of galaxy formation
simulations, analogous to the comparison of X-ray cluster formation carried
out in this paper is clearly desirable, but would be much more complex to
implement in practice. The level of agreement among the techniques
currently in use for studies of galaxy formation remains an open question.
Perhaps a simpler, but instructive, next step might be a comparison of
simulations of high redshift gas clouds (the ``Lyman-$\alpha$ forest''
clouds), in which gas cooling is less important.

Based on the overall consistency of the simulations discussed in this
paper, we can draw a number of conclusions regarding the properties of the
simulated cluster which we expect to be typical of near equilibrium,
massive clusters in a CDM universe. Some of these conclusions mirror those
found in earlier work (eg. Evrard 1990, White \etal 1993, Bryan \etal 1994,
Cen \& Ostriker 1994, Kang \etal 1994, Navarro, Frenk
\& White 1995, Anninos \& Norman 1996, Bartelman \& Steinmetz 1996.)

\noindent 1) The dark matter distribution in our simulated cluster 
is elongated along the direction of the dominant large filament along which
subclumps were accreted onto the cluster. The final gas distribution is
rounder than the dark matter distribution and, as a result, the direction
of the filament is difficult to identify in an X-ray image.

\noindent 2) The dark matter density profile in the cluster is well fit by
the analytic form proposed by Navarro, Frenk and White (1995), all the way
from 10 kpc to 10 Mpc. This form has a radial dependence close to $r^{-1}$
in the inner regions, steepens to $r^{-2}$ at intermediate radii and falls
off like $r^{-3}$ in the outer parts. The corresponding velocity dispersion
profile rises from the center outwards, has a broad maxium around 100-500
kpc, and declines in the outer parts.

\noindent 3) Although the gas is close to hydrostatic equilibrium
throughout the cluster, mergers disturb both the gravitational potential
and the dynamical state of the gas. Bulk motions make a small but
significant contribution to the support of the gas: the kinetic energy in
bulk gas motions is about 15\% of the thermal energy of the gas. The
quantity $\beta=\mu m_p\sigma_{DM}^2/3kT$ has a mean (averaged over all the
simulations except the lowest resolution one) of 1.15 with an rms scatter
of only 0.05.

\noindent 4) The radial density profile of the gas in the highest
resolution simulations develops a ``core radius", ie. a region in which the
slope of the profile flattens rapidly. This change of slope occurs at $r
\lsim 250$ kpc and, inside this radius, the gas density profile is
significantly flatter than the dark matter density profile. The gas
fraction within a given radius therefore rises from the center outwards
and, at the virial radius, the mean over all the simulations is 0.92 of the
global value, with a fractional rms scatter of only $\pm 0.065$. The
temperature distribution in the inner parts has an approximately flat
profile and, beyond a few hundred kpc, it begins to decline so that at the
virial radius, the temperature is $\sim 30$\% of the central value.

\noindent 5) A reliable estimate of the cluster X-ray luminosity requires 
resolving the radial range 200-500 kpc, or 5\%-20\% of the virial radius,
where the X-ray luminosity per logarithmic interval in radius peaks. Even
when this is possible, the strong sensitivity of X-ray luminosity to local
variations in gas density leads to a spread of a factor of $\sim 2$ in the
predicted X-ray luminosity. These variations are due to a variety of
numerical effects, and a factor of 2 uncertainty is a realistic estimate of
the accuracy with which cluster X-ray luminosities can be predicted with
the present generation of techniques and computing resources.

\acknowledgements 

We would like to thank Shaun Cole for generating the initial conditions for
the simulation, and Nigel Metcalfe for assistance in producing the
images. We acknowledge the hospitality of the Institute for Theoretical
Physics of the University of California, Santa Barbara, where this project
was initiated. This work was supported in part by the following grants:
PPARC Rolling Grant for ``Extragalactic Astronomy and Cosmology at
Durham''; EU TMR network for ``Galaxy formation and evolution;'' NATO
Collaborative Research Grant CRG 920182; NASA grants NAG5-2759, NAG5-2790,
NAG5-2882, and Long Term Space Astrophysics grant NAGW-3152; NSF grant
AST93-18185; DE-FG02-95ER40893; DGICyT of Spain project PB90-0182; NSERC
(Canada); and the CIES \& CNRS (France). Computing support provided by the
RZG Computing Center at Garching, EPCC at Edinburgh and the National Center
for Supercomputer applications is acknowledged. CSF acknowledges a PPARC
Senior Research Fellowship.

{}


\clearpage

\begin{figure}
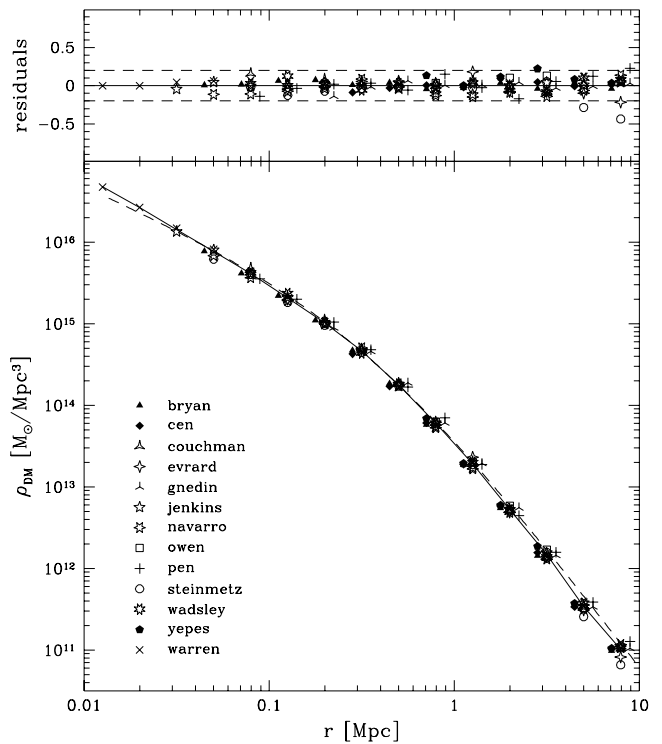

\figcaption{Projected dark matter density at $z=0$. The
images, covering the inner 8 Mpc of each simulation cube, have been
smoothed using the standard Gaussian filter of 250 kpc half-width described
in the text. Wadsley's simulation, not shown here or in Figure~2, has a
similar appearance to Couchman's. }
\end{figure} 

\begin{figure}
\caption{Projected dark matter density at $z=0.5$. The images,
covering the inner 8 Mpc of each simulation cube, have been smoothed using 
the standard Gaussian filter of 250 kpc half-width described in the text.}
\end{figure} 

\begin{figure}
\caption{Projected gas density at $z=0$. The images,
covering the inner 8 Mpc of each simulation cube, have been smoothed using 
the standard Gaussian filter of 250 kpc half-width described in the text.}
\end{figure} 

\begin{figure}
\caption{Projected gas density at $z=0.5$. The images,
covering the inner 8 Mpc of each simulation cube, have been smoothed using 
the standard Gaussian filter of 250 kpc half-width described in the text.}
\end{figure} 

\begin{figure}
\caption{Integrated, mass-weighted gas temperature at $z=0$. The images,
covering the inner 8 Mpc of each simulation cube, have been smoothed using
the standard Gaussian filter of 250 kpc half-width described in the
text. (The roughly circular ``cut-out'' regions seen in the outer parts of
this and the next figure are associated with cool, infalling clumps of size
comparable to the resolution of the smoothed image (see Figures~3 and~4);
the edges are enhanced by the choice of color table.)}
\end{figure} 

\begin{figure}
\caption{Integrated, mass-weighted gas temperature at
$z=0.5$. The images, covering the inner 8 Mpc of each simulation cube, have 
been smoothed using the standard Gaussian filter of 250 kpc half-width
described in the text.}
\end{figure} 

\begin{figure}
\caption{Projected X-ray luminosity at $z=0$. The images,
covering the inner 8 Mpc of each simulation cube, have been smoothed with the
filter chosen by each author to best portray the results of each simulation 
(see Table~1).}
\end{figure} 

\begin{figure}
\caption{Projected dark matter density at $z=0$. The images,
covering the inner 8 Mpc of each simulation cube, have been smoothed with the
filter chosen by each author to best portray the results of each simulation 
(ee Table~1).}
\end{figure} 

\clearpage
%
%
{\centerline{\bf Table 1: Simulation details}} 
\noindent (1) Name of principal simulator and reference to code description; 
(2) name of code; (3) $h$: spatial resolution for gas at cluster center
(cell-size or smoothing length of SPH kernel); (4) $M_{dm}$: dark matter
particle mass; (5) $M_{gas}$: gas mass resolution for Lagrangian codes
(particle mass or typical cell mass); (6) $\epsilon$: effective
gravitational force resolution (where appropriate this is the
gravitational softening length, defined as the effective softening of a
Plummer law fit to the actual gravitational force); (7) $\Delta t_{min}$:
minimum timestep; (8) $\Delta t_{ave}$: average timestep; (9) $N_{res}$:
number of gas resolution elements within the cluster (number of cells or
number of SPH gas particles); (10) image smoothing; (11) computer used;
(12) $T_{cpu}$: cpu processor hours; (13) Mem: memory required; (14) Date
of simulation.
\vskip 1cm
\begin{sideways}
\centering
\scriptsize
\begin{tabular}{|l|c|c|c|c|c|c|c|c|c|c|c|c|c|}
\hline
Name & code & $h$  & $M_{dm}$ & $M_{gas}$ & $\epsilon$ & $\Delta t_{min}$ &
$\Delta t_{ave}$ & $N_{res}$ & Image & Computer  & $T_{cpu}$ & Mem & Date \\
(Ref.)& & kpc & $10^{10}$\Mo & $10^9$ \Mo & kpc & & & & Smooth. & & 
hrs& Mbyte & \\
\hline\hline 
Bryan & SAMR & 15  & 0.78 &-- & 30 & 1/4000& 1/1000& $3.3\times 10^5$ &
Adaptive &SGI Pow. Ch.  & 200 & 500 & 3/96\\
(BN98) &&&&&&&&&$2h$&&&&\\
Cen   & TVD  & 125 & 0.10 &-- & 312 & 1/660& 1/660 & 36751 & Gauss &IBM
SP2  & 5312& 4400 & 1/96 \\
(ROKC93)&&&&&&&&&200 kpc  &&&& \\
Couchman & Hydra & 40 & 6.25 & 6.94 & 40 & 1/2500 & 1/1826 & 15291 
& Adapative& DECalpha & 77.3 & 95& 12/95 \\
(CTP95)&&&&&&&&&$1h$ &250 MHz&&& \\
Evrard & P$^3$M-SPH & 53 & 6.25 & 6.94 & 75 & 1/4000 & 1/4000 & 15571 & 
                Adaptive    & HP735 & 320 & 16.5 & 1/96 \\
(E88)&&&&&&&&& $3.5h$          &       &     & &        \\
Gnedin & SLH-P$^3$M& 100 & 6.25 & 6.94 & 100 & 1/4096 & 1/2315 & $1.5
\times 10^5$& Gauss & SGI Pow. Ch. & 136 & 90 & 9/97 \\
(G95)&&&&&&&&&100 kpc  &&&& \\ 
Jenkins & Par. Hydra & 20 & 0.78 & 0.87 & 20 & 1/20000& 1/4489 & $2.5 \times
10^5$ & Adaptive & Cray-T3D & 5000 & 512 & 4/96 \\
(PC97)&&&&&&&&&&&&& \\
Navarro & Grape+SPH& 30 & 6.18 & 6.87& 30 & 1/26074 & 1/651 & 13700 &
Gauss   & Sparc10    &120 & 75 & 4/96 \\
(NW93)  &            &    &      &     &    &         &       &       &
30 kpc  & + GRAPE-3AF &  &    &    \\
Owen& ASPH& 300 & 50 & 55.5 & 250 & 1/133713& 1/980& 1691 &Adaptive & Cray-YMP/4E 
& 40 & 106 & 3/96  \\ 
(OVSM98)&&&&&&&&&&&&&\\
Pen& MMH& 50 & 0.10 & 0.87 & 45 & 1/3523 & 1/1630 & 88953 &Adaptive & SGI Pow. Ch. &
480 & 900 & 4/96\\
(P98)&&&&&&&&&&&&&\\
Steinmetz & GrapeSPH & 50 & 6.25 & 6.94 & 25 & 1/7267 & 1/6500& 14876 &
Adaptive & Sparc10 & 28 & 22 & 4/96 \\ 
(S96)&            &    &      &     &    &         &       &       &
     & + GRAPE-3AF  &    &  &  \\
Wadsley &P$^3$MG-SPH&33.7& 6.25 & 6.94 & 24 & 1/4496 & 1/1630& 15918 &Adaptive &
Dec-Alpha EV5& 119& 100 & 7/96\\
(WB97)&&&&&&&&&$h\ge40$kpc&&&& \\
Warren & Tree& 5 & 0.11 & -& 5 & 1/2550 & 1/2550 & - & Adaptive & Intel-Delta& 15360 &
1000& 4/96\\
(WS95)&&&&&&&&&&&&& \\
Yepes &PM-FCT & 400 & 0.45 &-&960& 1/15000 & 1/6364 & 1024& None & Cray-YMP & 350&
480 & 11/95\\ 
(KKK92)&&&&&&&&&&&&& \\
\hline
\end{tabular}
\end{sideways}


\begin{thebibliography}{}

\bibitem{Ann96}Anninos, P. \& Norman, M.L. 1996, ApJ, 459, 12

\bibitem{Bal95} Balsara, D. S. 1995, J. Comput. Phys., 121, 357

\bibitem{BBKS} Bardeen, J.M., Bond, J.R., Kaiser, N. \& Szalay, A.S. 1986, 
ApJ, 304, 15

\bibitem{Bart96} Bartelmann, M. \& Steinmetz. M. 1996, MNRAS, 283, 421  

\bibitem{Bert95} Bertschinger, E. 1985, ApJ (Suppl.), 58, 39 

\bibitem{Berg89} Berger, M.J. \& Colella, P. 1989, J. Comput. Phys, 82, 64

\bibitem{Bo96} Bode, P.W., Xu, G., \& Cen, R. 1996, in Supercomputing 96, 
``http://dept.physics.upenn.edu/bode/SC96/INDEX.HTM

\bibitem{Bond96} Bond, J.R. \& Myers, S. 1996, ApJ (Suppl.), 103, 1

\bibitem{Bor71} Boris, J.P. 1971, in {\it Proceedings of the Seminar Course of
   Computing as a Language of Physics}, Intr Centre for Theoretical
   Physics, Trieste, Italy

\bibitem{Bor73} Boris, J.P. \& Book, D.L. 1973, J. Comp. Phys., 11, 38

\bibitem{Bor76} Boris, J.P. \& Book, D.L. 1976, J. Comp. Phys.,  20, 397

\bibitem{Bryan96} Bryan, G.L. Cen, R., Norman, M.L., Ostriker, J.P. \& Stone, J.M. 1994
ApJ, 428, 405

\bibitem{Bryan95} Bryan, G.L., Norman, M.L., Stone J.M., Cen, R. \& Ostriker,
J.P. 1995, Comput. Phys. Comm., 89, 149

\bibitem{Bryan95a} Bryan, G.L. \& Norman, M.L. 1995 BAAS, 187, 9504 

\bibitem{Bryan98} Bryan, G.L. \& Norman, M.L. 1998, in preparation (BN98) 

\bibitem{Cen92} Cen, R. 1992, ApJS, 78, 341. 

\bibitem{Cen90} Cen, R. Liu, F., Jameson, A. \& Ostriker J.P. 1990, ApJ Letters, 362,
L41


\bibitem{Cen94} Cen, R. \& Ostriker J.P. 1994, ApJ 429, 4  

\bibitem{Couch96} Couchman, H.M.P., Pearce, F.R. \& Thomas, P.A. 1996, astro-ph/9603116 

\bibitem{Couch95} Couchman, H.M.P., Thomas, P.A. \& Pearce, F.R. 1995, ApJ, 452, 797 (CTP
95)

\bibitem{Efs92} Efstathiou, G., Bond, R.J. \& White, S.D.M. 1992, MNRAS, 258, 1p 

\bibitem{Efs81} Efstathiou, G. \& Eastwood, J. W. 1981, MNRAS, 194, 503 

\bibitem{Evr88} Evrard, A.E. 1988, MNRAS, 235, 911 (E88) 

\bibitem{Evr90} Evrard, A.E. 1990, ApJ, 363, 349 

\bibitem{Ging77} Gingold, R. A \& Monaghan, J.J. 1977, MNRAS, 181, 375

\bibitem{Gned95} Gnedin, N.Y. 1995, ApJ (Suppl.), 97, 231 (G95) 

\bibitem{Gned96} Gnedin, N. Y. \& Bertschinger, E. 1996, ApJ, 470, 115 

\bibitem{Hart83} Harten, A. 1983, J. Comp. Phys., 49, 357

\bibitem{Hoff91} Hoffman, Y. \& Ribak, E. 1991, ApJ (Lett), 380, L5 

\bibitem{Huss98} Huss, A., Jain, B. \& Steinmetz, M., 1998, MNRAS, in press 

\bibitem{Kang94} Kang, H., Cen, R., Ostriker, J.P., \& Ryu, D. 1994a, ApJ, 428, 1 

\bibitem{Kang94b} Kang, H., Ostriker, J.P., Cen, R., Ryu, D., Hernquist, L., Evrard, A.E., 
Bryan, G., \& Norman, M.L., 1994b, ApJ, 430, 83

\bibitem{Kat90} Kates, R.E., Kotok, N. \& Klypin, A.A. 1990, A\&A 243, 295 

\bibitem{Kly92} Klypin, A.A., Kates, R.E. \& Khokhlov, A. 1992, in 'New Insights into the
Universe', eds. V. Mart\'{\i}nez, M. Portilla, \& D. Saez, Lecture Notes in
Physics, Springer-Verlag, 171 (KKK92)

\bibitem{Luc77} Lucy, L. 1977, AJ, 82, 1013

\bibitem{Mohr97} Mohr, J.J. \& Evrard, A.E. 1997, ApJ, 491, 38 
 

\bibitem{NFW95} Navarro, J.F., Frenk, C.S. \& White, S.D.M. 1995, MNRAS, 275, 720 

\bibitem{NFW97} Navarro, J.F., Frenk, C.S. \& White, S.D.M. 1997, ApJ, 490, 493 

\bibitem{NW93} Navarro, J.F., \& White, S.D.M. 1993, MNRAS, 267, 401 (NW93) 

\bibitem{Or86} Oran, E.S. \& Boris, J.P. 1986, {\it Numerical Simulation of
Reactive Flow}, Elsevier, New York

\bibitem{Ow98} Owen, J. M., Villumsen, J. V., Shapiro, P. R., \& Martel,
H. 1998, ApJS, 116, 155 (OVSM98)

\bibitem{Pea97} Pearce, F.R. \& Couchman, H. M. P. 1997, New Astr, 2, 411 (PC 97)

\bibitem{Pea96} Pearce, F.R., Couchman, H.M.P., Jenkins, A.R., \& Thomas, P.A. 1995,
in Dynamic Load Balancing on MPP systems

\bibitem{Pen95} Pen, U.-L. 1995, ApJ (Suppl.), 100, 269 

\bibitem{Pen98} Pen, U.-L. 1998, ApJ (Suppl.), 115, 19 (P98) 

\bibitem{Po85} Porter, D., 1985. PhD thesis UC Berkeley 

\bibitem{Ryu93} Ryu, D., Ostriker, J. P., Kang, H., \& Cen, R. 1993, 414, 1 (ROKC93)


\bibitem{Shap96} Shapiro, P.R., Martel, H., Villumsen, J.V. \& Owen, J.M. 1996,
ApJ (Suppl.), 103, 269

\bibitem{Stei96} Steinmetz, M., 1996, MNRAS, 278, 1005 (S96) 

\bibitem{Stei93} Steinmetz, M., M\"uller, E., 1993, A\&A, 268, 391 

\bibitem{Sugi90} Sugimoto, D., Chikada, Y., Makino, J., Ito, T., Ebisuzaki, T. \& Umemura, 
M. 1990, Nature, 345, 33 

\bibitem{Wad97} Wadsley, J.W. \& Bond, J.R. 1997, Proc. 12th Kingston
Conf., Halifax, Oct. 1996, 332, ed. D. Clarke \& M. West (PASP),
astro-ph/9612148 (WB97) 

\bibitem{War93} Warren, M.S. \& Salmon, J.K. 1993, ``A parallel hashed oct-tree N-body
algorithm,'' in Supercomputing '93, pages 12-21, Los Alamitos,
1993. IEEE Comp. Soc. http://qso.lanl.gov/papers/sc93/reprint.ps 

\bibitem{War95} Warren, M.S.  \& Salmon, J.K. 1995,  Computer Physics Communications, 
87, 266. (WS95) 

\bibitem{Whi96} White, S.D.M. 1996, in Cosmology and Large-scale structure, Elsevier, 
Dordrecht, eds. Schaefer, R, Silk, J., Spiro, M. and Zinn-Justin, J. 

\bibitem{Whi93} White, S.D.M., Navarro, J.F., Evrard, A.E. \& Frenk, C.S. 1993, Nature,
366, 429 

\bibitem{Yep95} Yepes, G., Kates, R., Klypin, A.  \& Khokhlov, A., 1995, 
Proceedings of the XVth Recontres des Moriond {\it ``Clustering in the
Universe''}. Ed. S. Maurogordato \etal pg. 209

\bibitem{Yep96} Yepes, G., Kates, R., Klypin, A.  \& Khokhlov, A., 1996.
Proceedings of the UIPM-ECN Conference {\it `` Mapping, Measuring and Modelling
the Universe''}.  Ed.  M.  J.  Pons, V.  Mart\'{\i}nez and P.  Coles.
PASP.  pg 125 

\end{thebibliography}
\end{document}